\documentclass[useAMS,usenatbib]{mnras}
\usepackage{graphicx}
\usepackage{amssymb}

\title[Spatial segregation and SF in nearby dSphs]{Spatial segregation impact
on star formation in nearby dwarf spheroidal galaxies
\thanks{Based on observations made with the NASA/ESA Hubble
Space Telescope, program GO-13442, with data archive at the Space
Telescope Science Institute. STScI is operated by the Association of 
Universities for Research in Astronomy, Inc. under NASA
contract NAS 5--26555.}
}
\author[L. N. Makarova and D. I. Makarov]{
L. N. Makarova$^{1}$\thanks{E-mail: lidia@sao.ru}, 
D. I. Makarov$^{1}$\\
$^{1}$Special Astrophysical Observatory, Nizhniy Arkhyz, 
Karachai-Cherkessia 369167, Russia
}
\begin{document}

\date{Accepted XXX. Received XXX; in original form XXX}

\pagerange{\pageref{firstpage}--\pageref{lastpage}} \pubyear{XXX}

\maketitle

\label{firstpage}

\begin{abstract}
Using our HST/ACS observations of the recently found isolated dwarf spheroidal
galaxies, we homogeneously measured their star formation histories. We 
determined star formation rate as a function of time, as well as age and 
metallicity of the stellar populations. All objects demonstrate complex star 
formation history, with a significant portion of stars formed 10­--13 Gyr ago. 
Nevertheless, stars of middle ages (1--­8 Gyr) are presented. In order to 
understand how the star formation parameters influence the evolution of dSphs, 
we also studied a sample of nearest dSphs in different environment: isolated 
($d < 2$ Mpc); beyond the Local Group virial radius (but within the Local Group 
zero velocity sphere); and the satellites of M\,31 located within the virial 
zone (300 kpc). Using archival HST/ACS observations, we measured their star
formation histories. A comparative analysis of the parameters obtained allow
us to distinguish a possible effect of the spatial segregation 
on the dSphs evolution scenario.
\end{abstract}


\section{Introduction}
A morphological type dependence of a galaxy on its distance to a central object
in galaxy groups and clusters is well known for a long time (the distance--morphology 
relation phenomenon). \citet{oemler1974} and \citet{dressler1980} have found 
a well defined relationship between local galaxy density and galaxy type in 
rich clusters. This relationship indicates an increasing of elliptical and SO galaxy 
population, and a corresponding decrease of spiral galaxy number with increasing 
density. Clusters rich in spirals, in turn, demonstrate an irregular distribution 
and low mass density, lack of concentration towards the centre and no signs 
of segregation by mass or morphological type of galaxies. Obvious signs of the 
same behaviour are noticeable in the galaxy groups closest to us. In our Local Group (LG), 
for example, there are clearly visible concentrations of dwarf spheroidal (dSph)
galaxies, the satellites of the Milky Way and Andromeda (M\,31) 
within their virial radii (see Fig.\,7 in the paper of \citet{mak2017} 
and references therein). This fact may serve as an indication that the main 
mechanism of gas loss and star formation quenching in dwarf galaxies 
of the Local Group is associated with the proximity of the central galaxy.

\begin{figure*}
\begin{tabular}{cc}
\includegraphics[width=7cm]{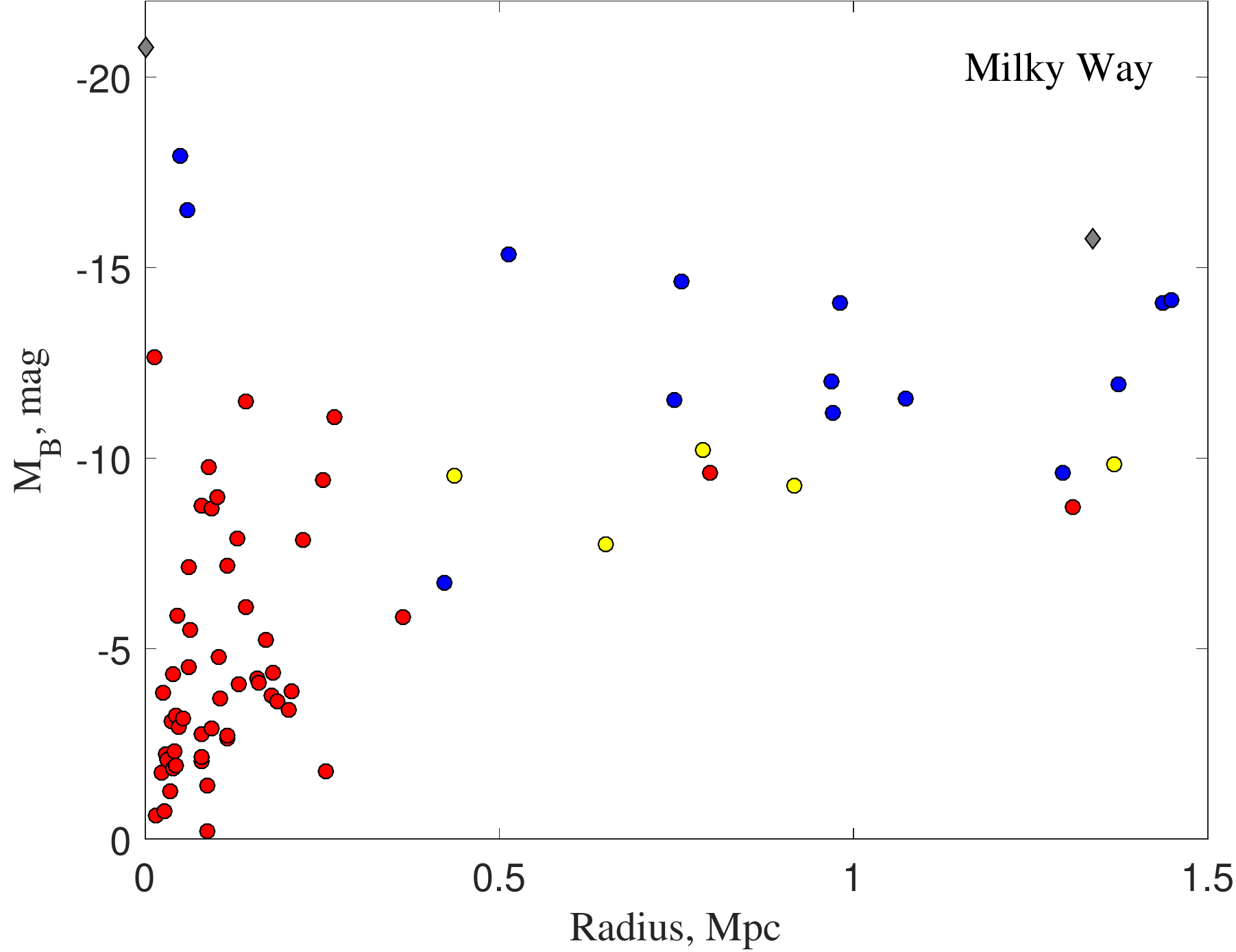} &
\includegraphics[width=7cm]{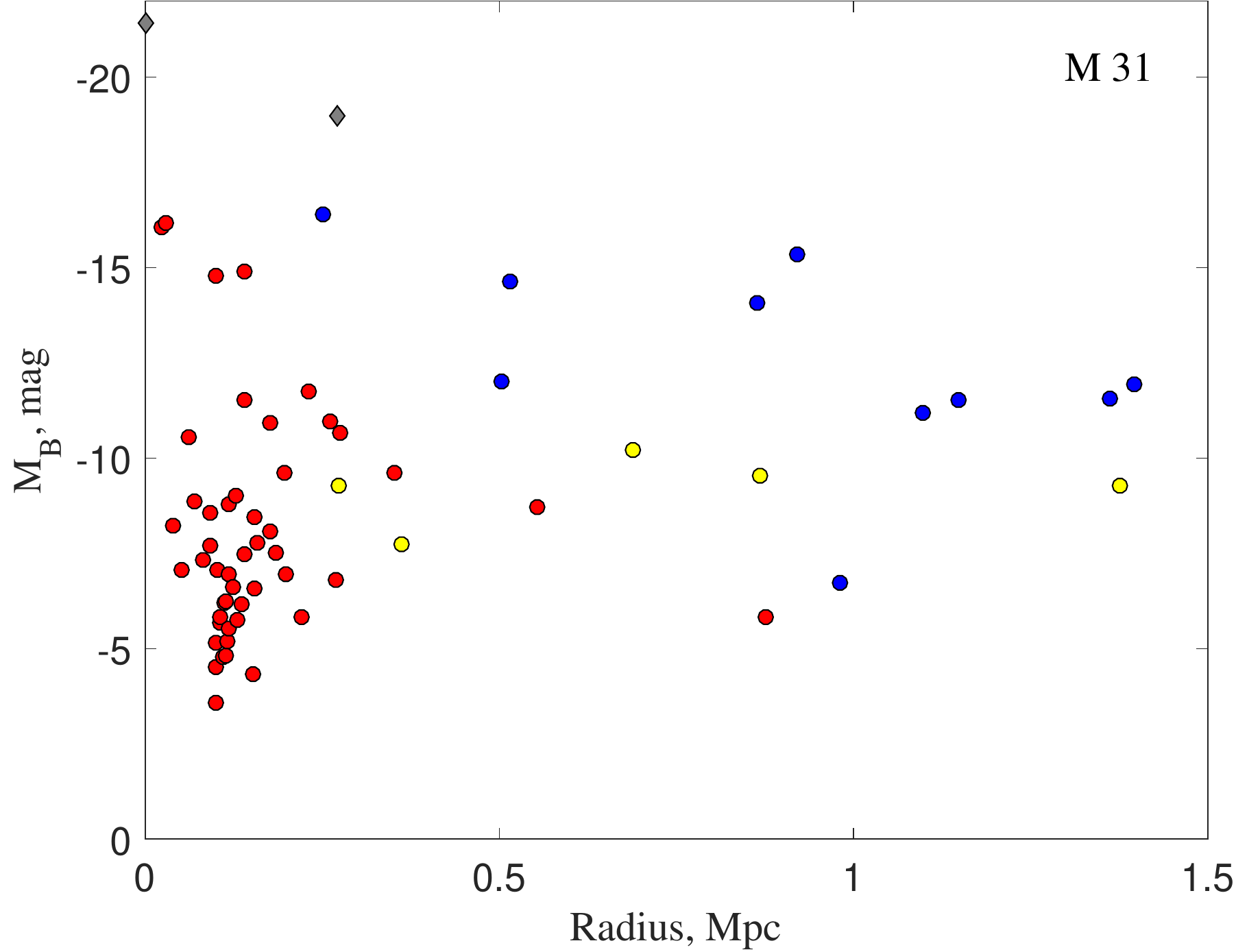} \\
\includegraphics[width=7cm]{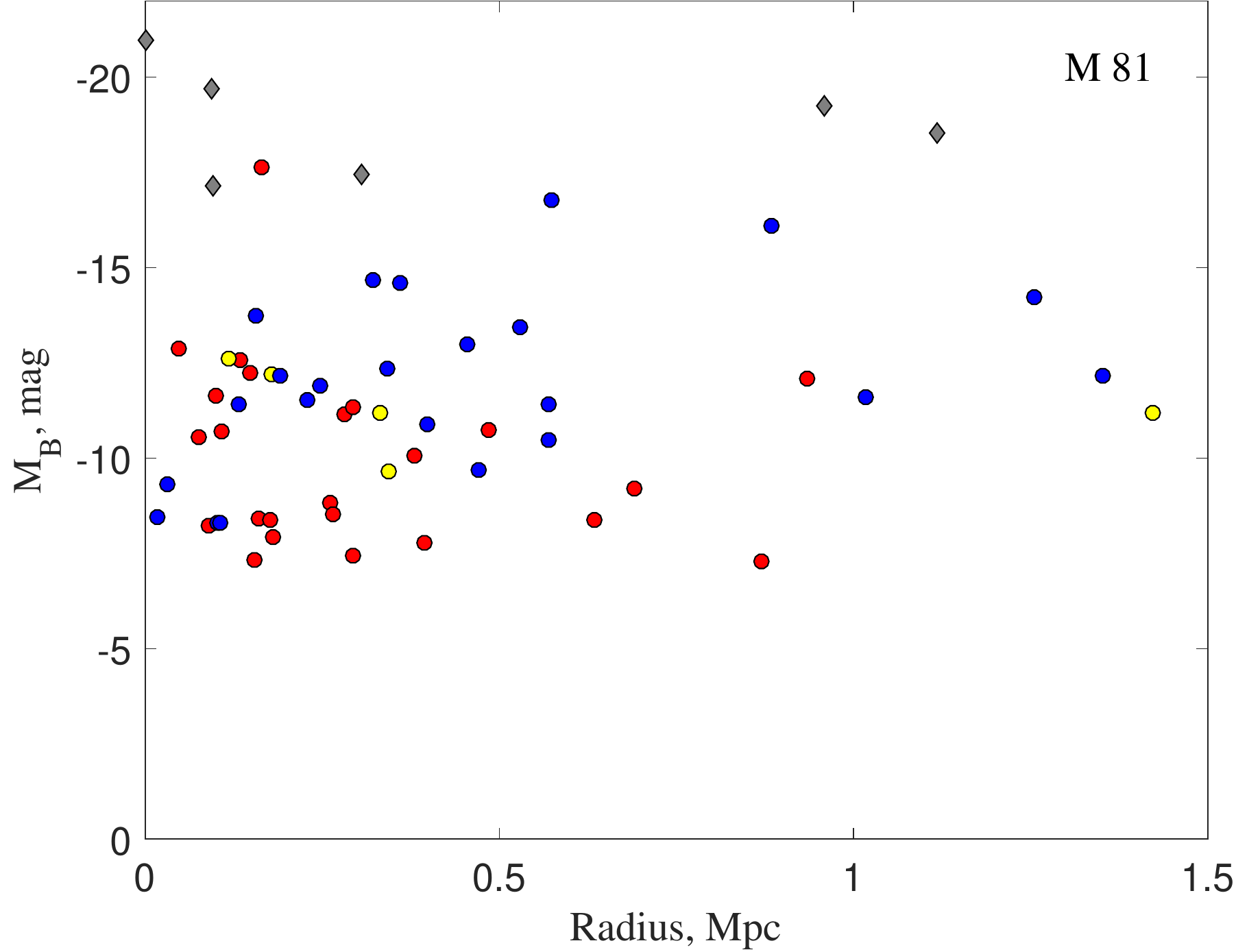} &
\includegraphics[width=7cm]{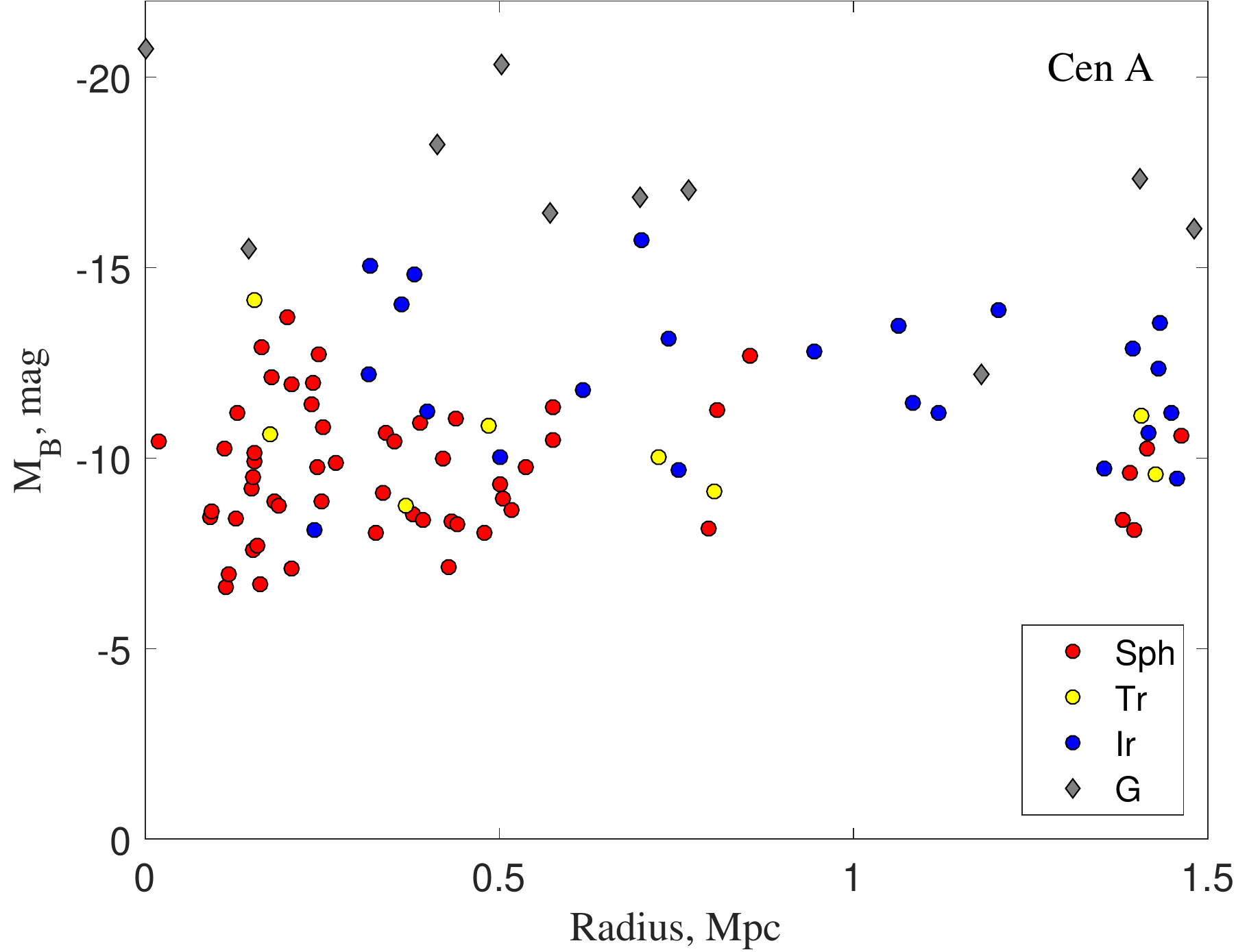} \\
\end{tabular}
\caption{Distribution of an absolute magnitude with the distance of the galaxy
from the central object in the group. This relation is constructed for dwarf 
satellite galaxies within 1.5 Mpc from the centre for the four nearest groups: 
subgroup of our Galaxy, Andromeda subgroup, M\,81 group and Cen A group. 
The coloured circles indicate the morphological type of a galaxy: red -- dwarf 
spheroidal; blue -- dwarf irregular; yellow -- dwarfs of transitional type; 
grey -- other morphological types. The data for this relations is taken from 
our Local Volume (LV) galaxy database.
}
\label{fig:morph}
\end{figure*}

Figure~\ref{fig:morph} shows the distribution of the absolute magnitudes of 
galaxies corrected for Galactic and internal extinction with the distance
of a galaxy from the central object in the group for the four nearest groups:
a subgroup of our Galaxy (Milky Way), the Andromeda (M\,31) subgroup, M\,81 group
and Centaurus A group, taking into account morphological type of galaxies. 
The concentration of dwarf spheroidal galaxy satellites within about 300 kpc 
from the centre of the group is clearly visible. The distance of about 300 kpc
roughly corresponds to the virial radius \citep{klypin2002}, 
and 1 Mpc is related to the zero-velocity sphere
radius \citep{karach2009}. 
Dwarf irregular galaxies and 
transitional type galaxies are distributed more or less evenly within the groups. 

At the same time, not all dwarf spheroidal satellites in the groups are located
within 300 kpc from the host galaxy (see Fig.~\ref{fig:morph}). There are 
individual dSphs located at a distance from $\sim 0.5$ to $1.5$ Mpc from 
the centre of the group, i.e. some of these systems may be actually spatially isolated. 
It is obvious that the gas loss and star formation quenching mechanisms in these 
galaxies should be different and not related to the interaction with the central 
galaxy of the group \citep{rocha2012, teyssier2012}.

One of the great advantages of the study of dwarf galaxies in nearby groups (within about 10--12 Mpc)
is a possibility to resolve these objects into individual stars, including red giant branch  
(RGB) and fainter stars. Size of such galaxies is usually small, and the structure 
is relatively simple, which simplify the task of detailed study of their
star formation histories (SFHs). 
Dwarf galaxies of the Local Group and its immediate environment (within 1--2 Mpc from us)
are especially promising. These objects
can be resolved into stars, including red clump (RC) of the red
giant branch at the Hubble Space Telescope (HST) with relatively small exposures.
Considerable efforts has been made to determine SFHs of dwarf galaxies in
our Milky Way Group and M\,31 group of galaxies. Only in the last few years
we can note a number of large studies of satellite dwarf spheroidal galaxies
(see, for example, \citet{skillman2017}, \citet{martin2017}, \citet{kirby2020}). 
Different samples presented in these researches
drive our attention to possible quenching time for the dSphs, role of reionization epoch
and the environmental influence (e.g. \citet{bullock2017}).

In this paper we present detailed stellar photometry and comparative star formation history
for a sample of 8 dwarf spheroidal galaxies of the Local Group. The galaxy images
were obtained at the HST with ACS (Advanced Camera for Surveys). 
We have selected objects located at different distances within the Local Group and its 
vicinity. Our aim is to provide uniform processing and stellar photometry of images, as well as
homogeneous star formation history measurements, to get the possibility of 
a comparative study of the dwarf spheroidal galaxy evolution and possible star
formation quenching mechanisms.

\section{Dwarf spheroidal galaxy sample selection}
\label{sec:sample}
Dwarf spheroidal galaxies located outside the zero velocity radius of the Local 
Group have come to the focus of our attention in recent years. We have discovered 
and studied a number of such objects, including the extremely isolated ones 
\citep{mak2012,kar2014,kar2015,mak2017}. Most of these galaxies were observed with
the HST/ACS in the framework of the programs 12546 and 13442. The next important 
step to study possible mechanisms of evolution and star formation quenching
of nearby dSphs is the selection of a test sample of such 
dwarf galaxies of the Local Group, located at different distances from 
the gravitational centre of the group, which have archival data from 
the HST and ACS/WFPC2 cameras. The data obtained with the same 
instruments, similar exposures, and processed uniformly, allow us to intercompare 
the details of measured star formation history of the dwarfs, without corrections 
assumed for various data analysis techniques and observational conditions.

At the same time, we cannot avoid in our analysis an important non-uniformity
within our sample caused by different distance to the objects under study.
With comparable photometric depth less luminous age-sensitive features are visible 
for closer galaxies. Detailed analysis and tests
by \citet{weisz2011} show systematic uncertainties in SFH caused by difficulties
in modelling such features like RGB, horizontal branch and red clump.
For deeper colour-magnitude diagrams ($M_V$ = +1, which correspond to the isolated
dwarfs from our sample) the authors report about 40 per cent overestimation of oldest star
formation. Good recovering of star formation history has shown for deepest cases,
which include MSTO ($M_V$ = +4), but all our objects are falls in previous photometric
limits ($M_V$ roughly between +1 and +3). Therefore, we would expect, that SFHs measured by us
has systematic uncertainties working in the similar way and they are not affect 
out conclusions.
\citet{weisz2011} also estimated the systematic uncertainties within the last 2 Gyr
to be better than 10 per cent at all photometric depths. This result is  
important for our discussion of quenching time and residual star formation in
the selected dSph galaxies.

From the Local Group list \citep{upgc2013} we have chosen dwarf 
galaxies with a Holmberg diameter $a_{26}\leq3$ arcmin. This size fits
the entire or most of the galaxy into the HST/ACS camera field ($3.4\times3.4$ arcmin). 
We have excluded from our consideration the objects 
located within 100 kpc from the Milky Way and Andromeda. Obviously
for these satellites the influence of the host galaxy dominates in their 
evolution and star formation, due to the high probability of tidal 
interaction, which can both sweep the gas out of the dwarf galaxy, thereby causing 
the star formation quenching, and also induce the formation of tidal dwarf galaxies
from filaments of perturbed intergalactic medium \citep {mak2002}.
The other criteria of our test sample selection simply reflect the availability
of HST/ACS exposures with appropriate filters and duration to complement our sample
of the isolated dwarf spheroidals (see the references in the beginning of this
section). A the same time it it better to restrict our sample by the satellites of
the same galaxy for the straightforward comparison of the selected objects
evolution.

Thus, 8 dwarf spheroidal (or transitional type) galaxies that are satellites of 
the neighbouring spiral galaxy M\,31, or isolated objects, fall into our sample.

The list of the galaxies and their general parameters are given in Table~\ref{tab:param}. 
These data are mainly taken from the LV catalogue~\footnote{\url{https://www.sao.ru/lv/lvgdb/}} \citep{upgc2013}. 
The first column of the table shows the name of the galaxy; 
(2) -- equatorial coordinates (J2000); (3) -- morphological type; 
(4) -- absolute magnitude in Johnson B filter; (5) - Holmberg diameter 
(at the level of $\mu_B = 26.5$ mag arcsec$^{-2}$ isophote), in angular 
minutes; (6) -- the distance in Mpc. The distances to isolated galaxies 
were measured by us using tip of the red giant branch method (TRGB) in the works 
mentioned above. 

The three subsample of the list are indicated in the Table~\ref{tab:param}: highly
isolated dwarfs KKR\,25, KK\,258 and KKs\,03; the dSphs within the zero velocity
radius, but outside the virial radius Tucana dwarf and And\,XVIII; the dSphs
within the virial radius of M\,31 And\,XXVIII, Cas\,dSph and And\,XXIX.

\begin{table}
\centering
\caption{General parameters of the sample galaxies}
\label{tab:param}
\begin{tabular}{lllrrr}
Name         & R.A.\,Decl        & Type   & $M_B$    & a26    & $D$    \\ 
             &  (J2000)          &       &  mag     & ' & Mpc   \\
\hline
KKR 25       & 161347.7+542215   & dSph  & $-$9.44   & 1.10  & 1.91  \\
KK 258       & 224043.8$-$304758 & dTr   & $-$10.51  & 1.70  & 2.24  \\
KKs 03       & 022443.5$-$733049 & dSph  & $-$10.72  & 2.45  & 2.00  \\
\multicolumn{6}{c}{\dotfill} \\
Tucana      & 224149.0$-$642512 & dTr   & $-$9.26   & 2.88  & 0.92  \\
And XVIII   & 000214.5+450520   & dSph  & $-$8.73   & 1.60  & 1.31   \\
\multicolumn{6}{c}{\dotfill} \\
And XXVIII  & 223241.2+311258   & dTr   & $-$7.72   & 1.70  & 0.65   \\
Cas dSph    & 232631.8+504032   & dSph  & $-$11.76  & 3.02  & 0.82   \\
And XXIX    & 235855.6+304520   & dSph  & $-7$.52   & 2.70  & 0.73   \\
\hline
\end{tabular}
\end{table}

\section{Stellar photometry}
\label{phot}
Observations of all studied dwarf galaxies (with the exception of KKR\,25) 
were made with HST/ACS. 
The observational data are listed in Table~\ref{tab:obs}, where the first column gives 
a galaxy name; (2) -- HST instrument used; (3) -- filters; (4) -- exposure time
 we used in our reduction; (5) -- total exposure time available for this 
observational program. We did not use all available data for Tucana, And XXVIII and Cas dSph,
as they are not make sense for our purposes (see the discussion on the data sample above 
in the Section~\ref{sec:sample});
(6) -- publication link for the SFH measurements used in the present work; (7) -- SFH determined
by other authors with HST/ACS or WFPC2 observations.

\begin{table*}
\centering
\caption{HST observations of the dwarfs}
\label{tab:obs}
\begin{tabular}{lcccclll}
Name       & Instrument  & Filter   & $t_{used}$  & $t_{all}$ & Project (PI) & Reference & Other references \\
\hline
KKR 25     & HST/WFPC2   & F606W    & 4800s   &     & 11986 (J. Dalcanton) & \citet{mak2012} & \citet{weisz2011} \\
           &             & F814W    & 9600s   &     & 11986 &   &  \\
KK 258     & HST/ACS     & F606W    &  900s   &     & 12546 (R. Tully)     & \citet{kar2014} & \\
           &             & F814W    &  900s   &     & 12546 &   &  \\
KKs 03     & HST/ACS     & F606W    & 1200s   &     & 13442 (R. Tully)     & \citet{kar2015} & \\
           &             & F814W    & 1200s   &     & 13442 &   &  \\
\multicolumn{8}{c}{\dotfill} \\
Tucana     & HST/ACS     & F475W    & 2160s   & 34560 & 10505 (C. Gallart)   & this work  & \citet{savino2019}   \\
           &             & F814W    & 1936s   & 30976 & 10505 &   &  \\
And XVIII  & HST/ACS     & F606W    & 1100s   &     & 13442 (R. Tully)     & \citet{mak2017} & \citet{weisz2019} \\
           &             & F814W    & 1100s   &     & 13442 &   &  \\
\multicolumn{8}{c}{\dotfill} \\
And XXVIII & HST/ACS     & F475W    & 2636s   & 26360 & 13739 (E. Skillman)  & this work & \citet{skillman2017}  \\
           &             & F814W    & 2088s   & 20880 & 13739 &  &   \\
Cas dSph   & HST/ACS     & F435W    & 4800s   & 38400 & 10430 (T. Armandroff) & this work & \citet{weisz2014}    \\
           &             & F555W    & 4800s   & 19200 & 10430 &   &  \\
And XXIX   & HST/ACS     & F606W    & 1118s   &       & 13699 (N. Martin) & this work  & \citet{weisz2019}   \\
           &             & F814W    & 1117s   &       & 13699 &   &  \\
\hline
\end{tabular}
\end{table*}

Initial data reduction of the images was made within the default HST pipeline. 
We use the DOLPHOT software package by \citet{dolphin2000}\footnote{\url{http://americano.dolphinsim.com/dolphot/}} 
for stellar photometry of 
resolved stars in all galaxies of the sample. The photometric reduction procedures 
were made according recipes in the DOLPHOT's User Guide. For further analysis, 
we selected only stars of good photometric quality: type=1, a signal-to-noise ratio 
(S/N) of at least five in both filters and $\vert sharp \vert \le 0.3$ .

We performed photometric tests with artificial stars for the most accurate estimation 
of photometric errors, incompleteness of photometry and blending of stars in the
crowded stellar fields. These tests were carried out for the sample objects using 
the same photometric procedures. A sufficient number of artificial stars 
(approximately ten times greater than the number of real measured ones) was 
generated in the respective range of stellar magnitudes and colour indices, 
so that the distribution of the restored magnitudes corresponded to the real one.

\section{Colour--magnitude diagrams}

\begin{figure*}
\begin{tabular}{ccc}
\includegraphics[height=6cm]{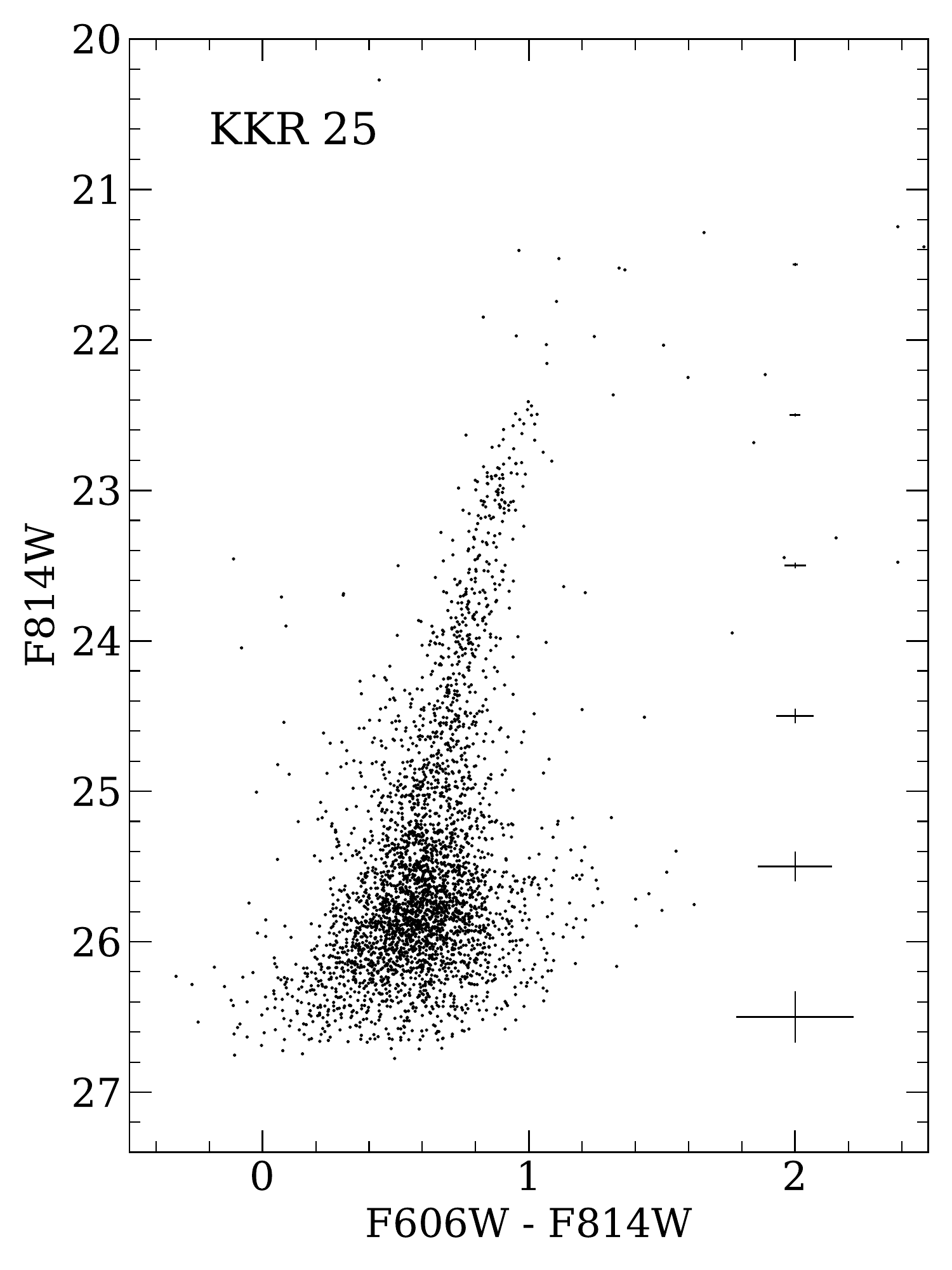} &
\includegraphics[height=6cm]{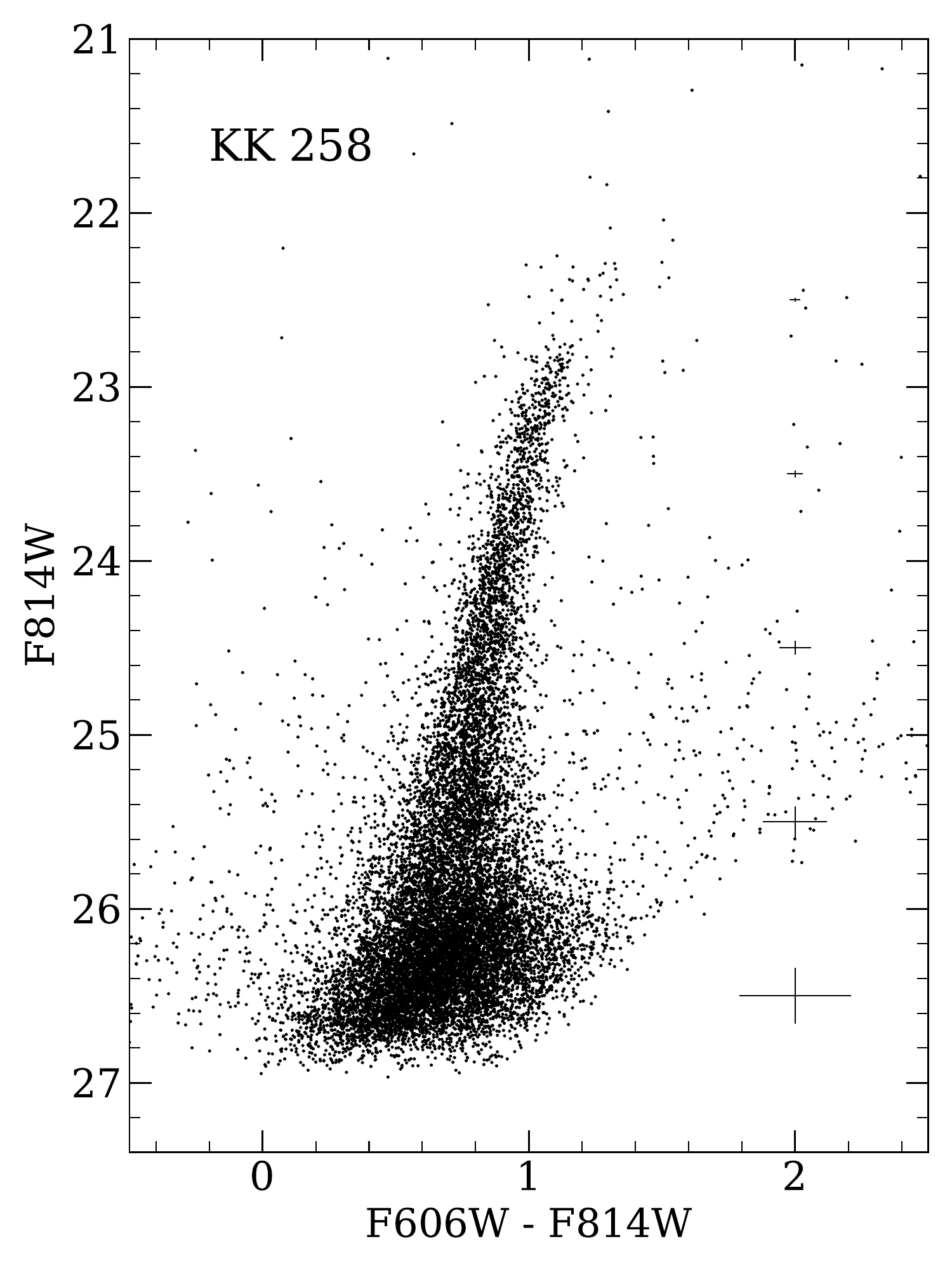} &
\includegraphics[height=6cm]{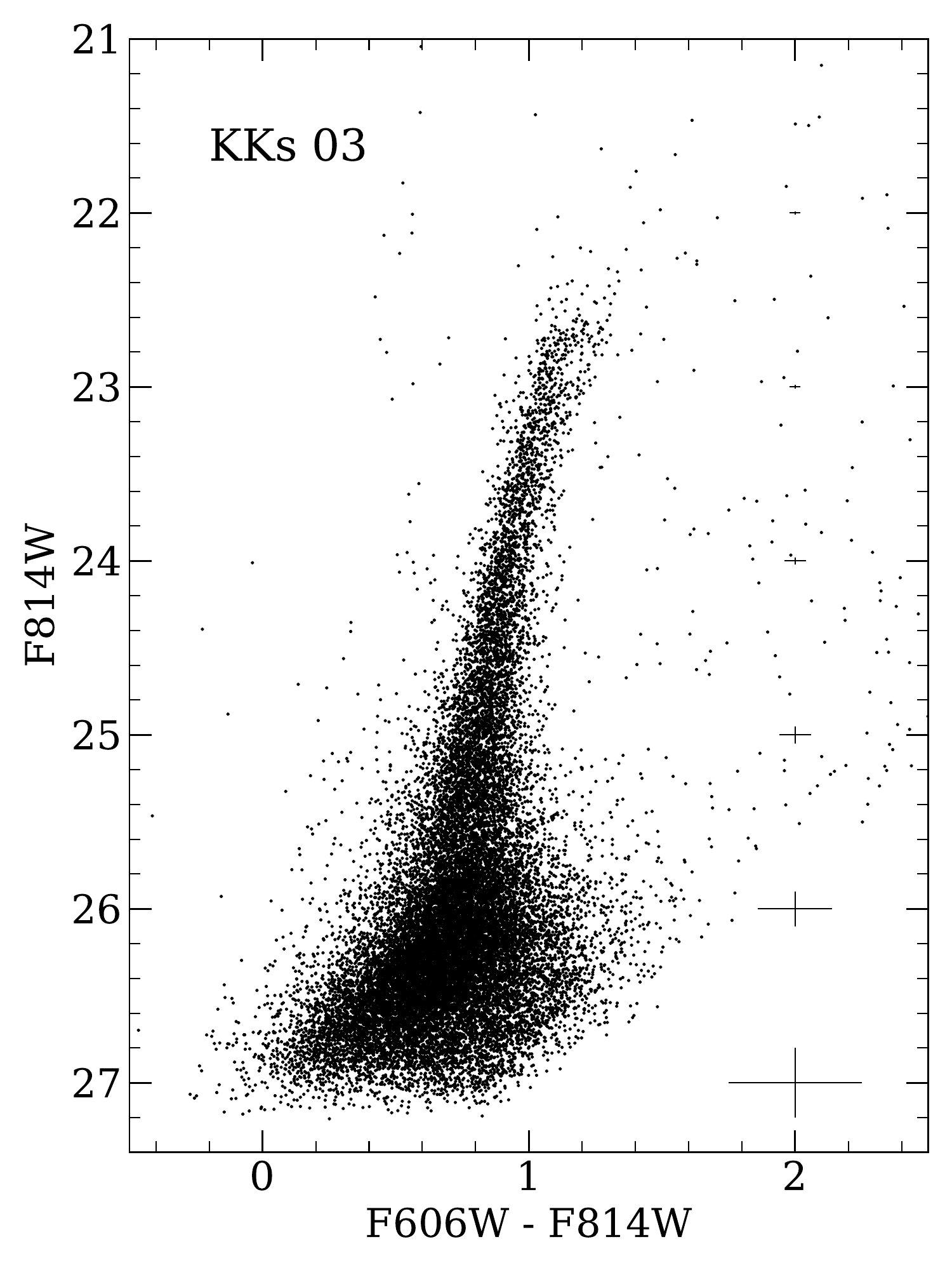} \\
\includegraphics[height=6cm]{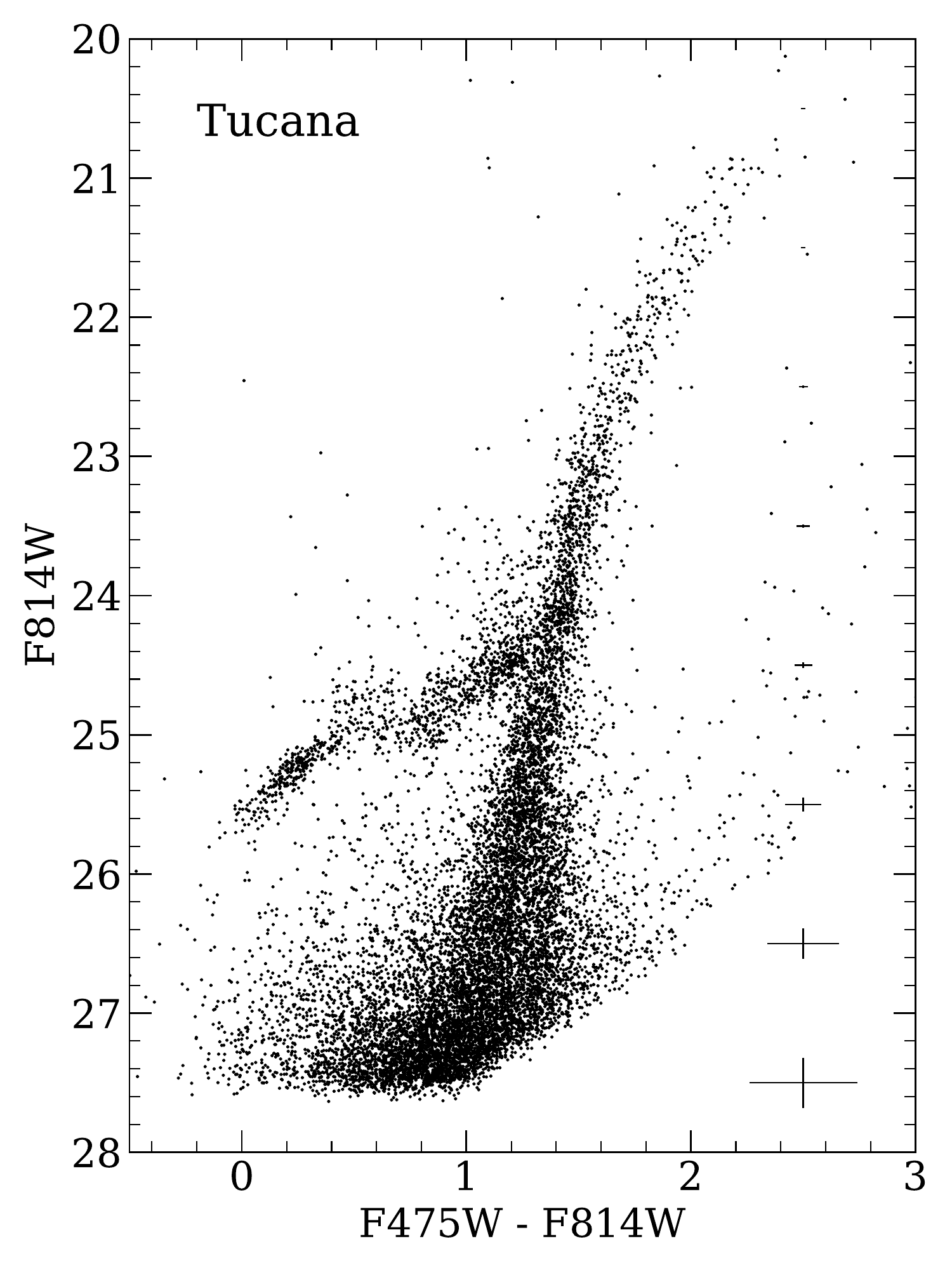} &
\includegraphics[height=6cm]{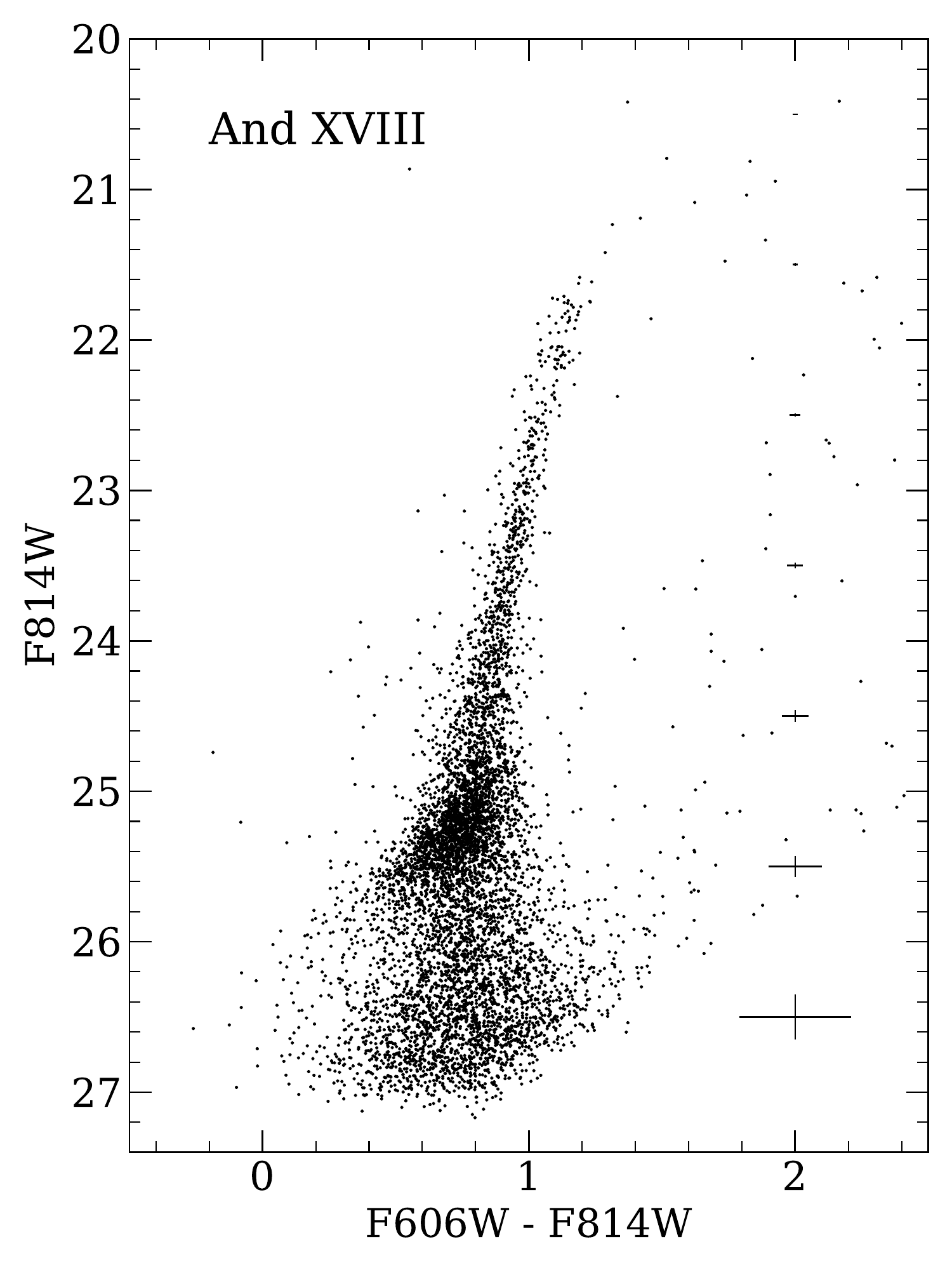} & \\
\includegraphics[height=6cm]{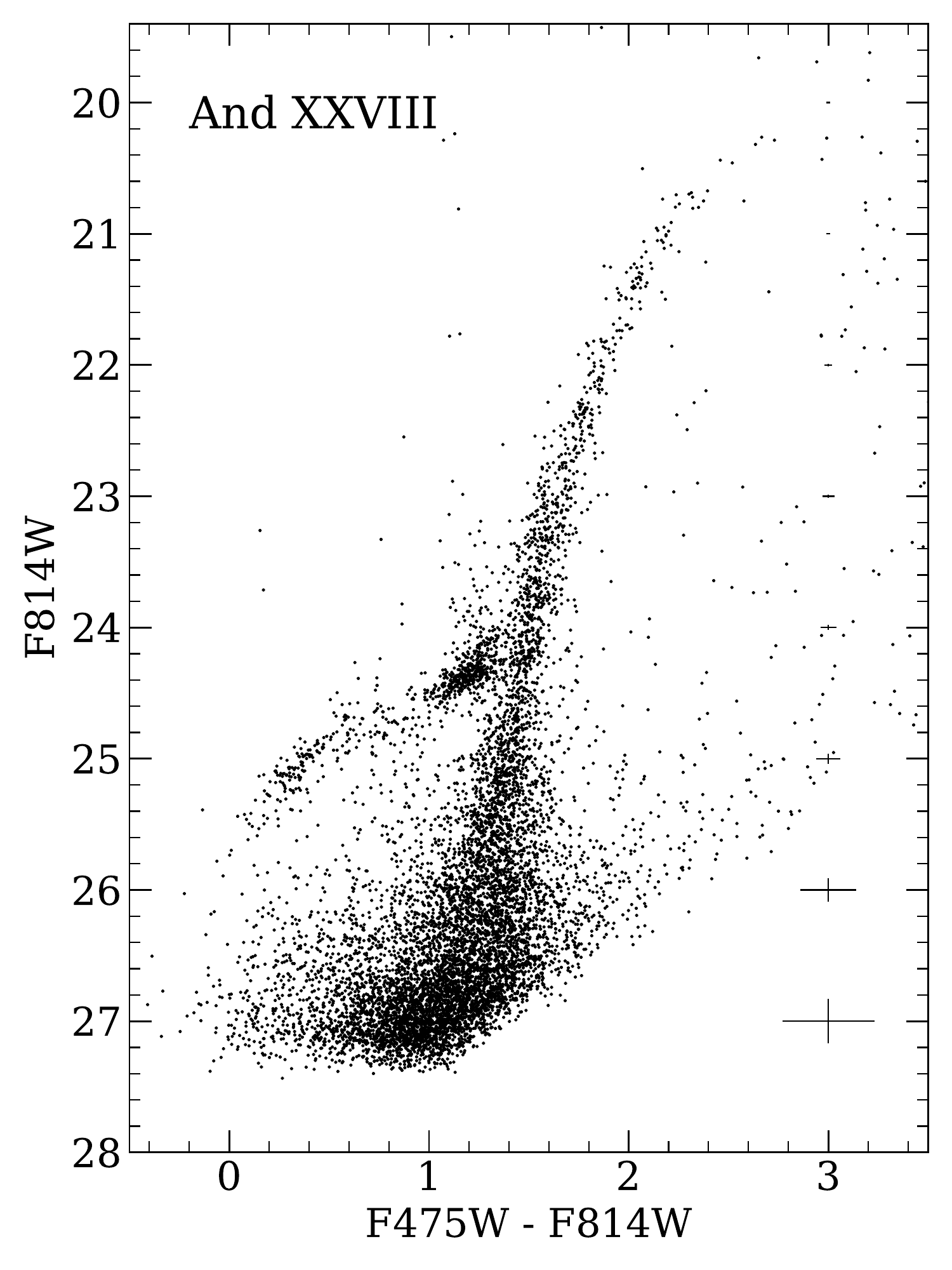} &
\includegraphics[height=6cm]{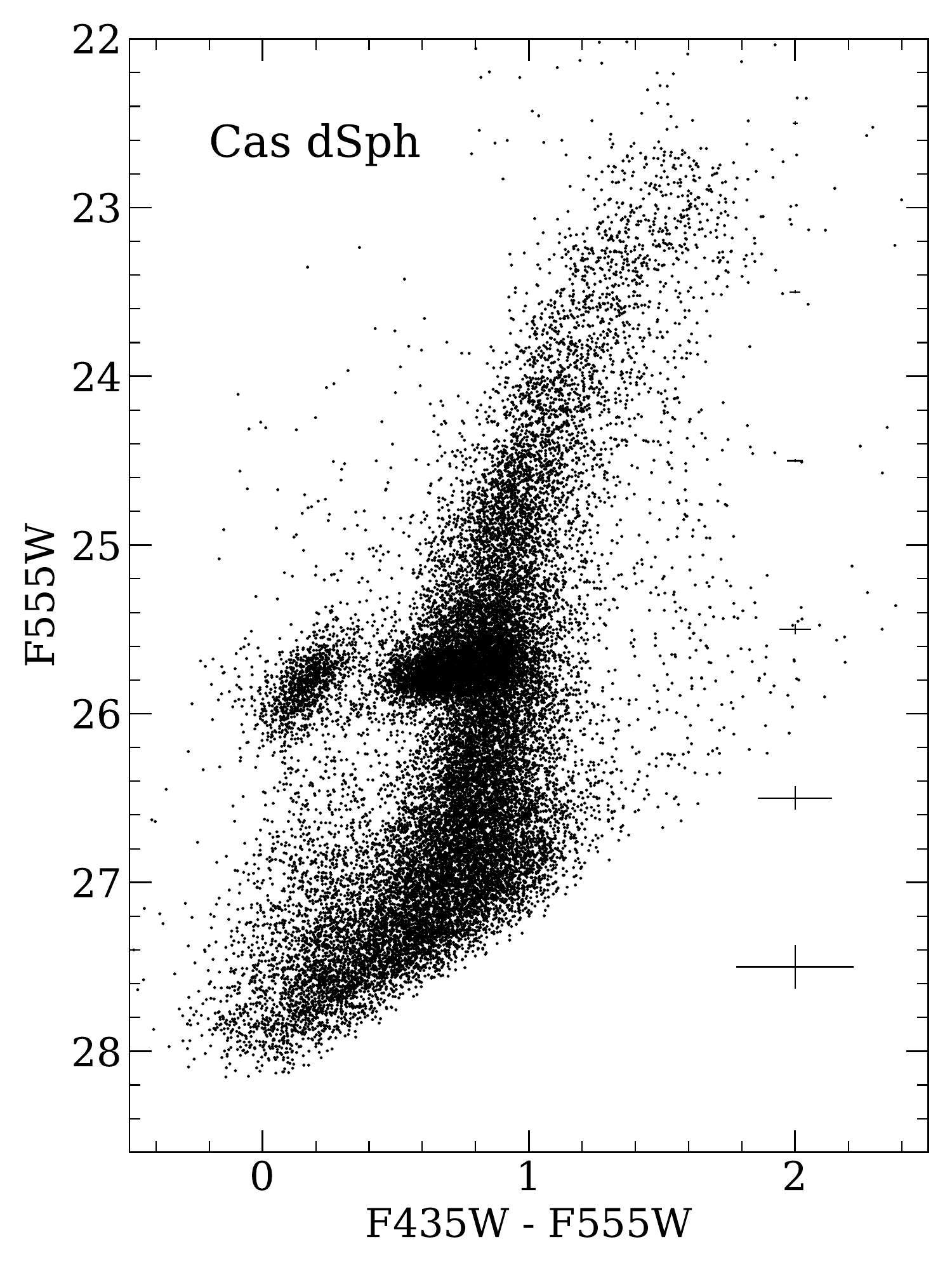} &
\includegraphics[height=6cm]{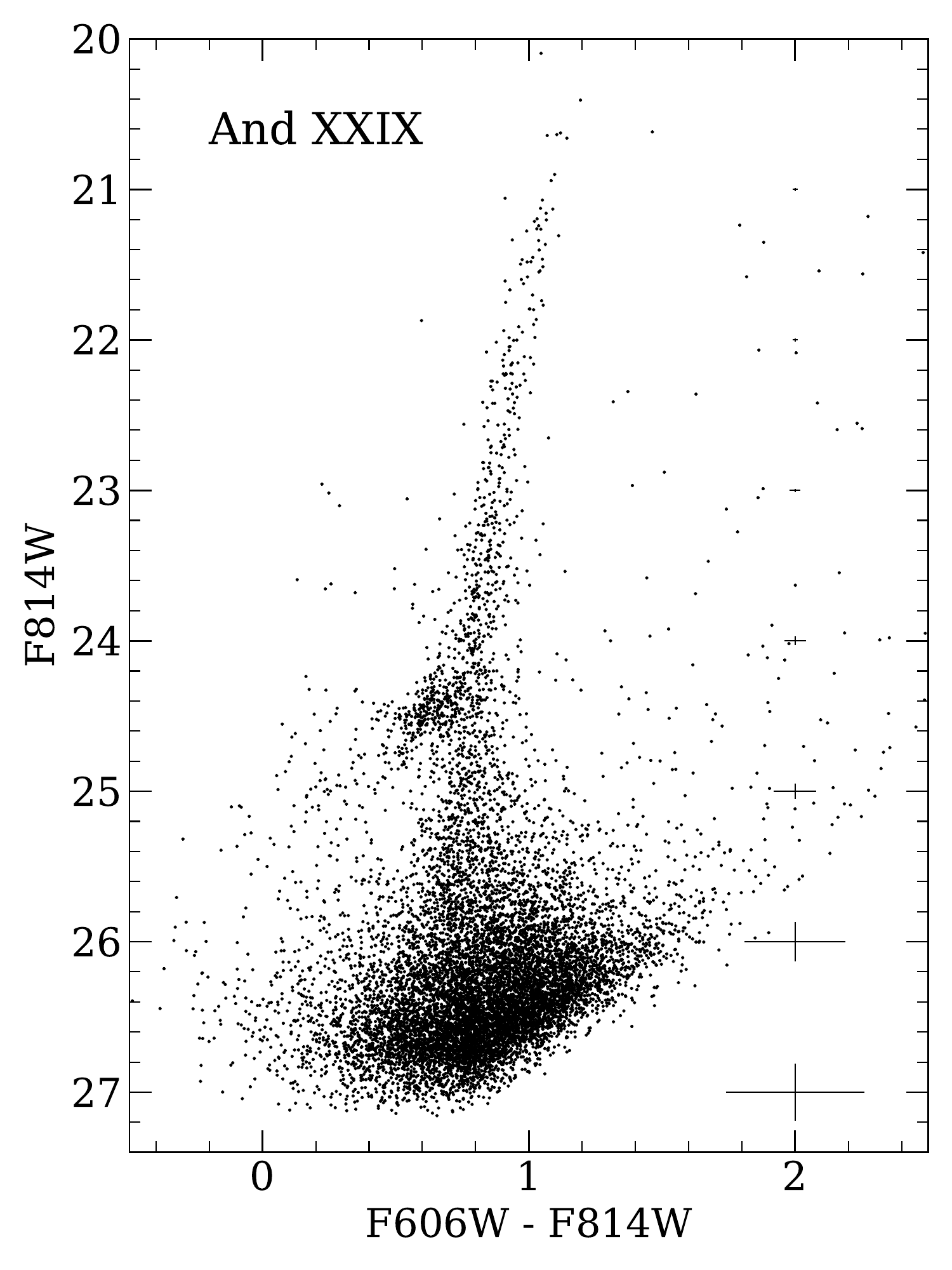} \\
\end{tabular}
\caption{Colour-magnitude diagrams of the dwarf galaxies of the sample.
Photometric error bars are shown at the right part of each CMD.
}
\label{fig:cmd}
\end{figure*}

The resulting colour-magnitude diagrams (CMDs) for our sample of galaxies are shown in 
Fig.~\ref{fig:cmd}. Photometric error bars are shown at the right part of each diagram.
The photometric limit for all dwarfs is nearly the same (about 27--27.5 mag in $F814W$ filter).
The relatively narrow red giant branch (RGB) forms the dominant part of the resolved stellar populations
of the studied dwarfs. Certain exception is the brightest galaxy in our sample, Cas\,dSph, 
demonstrating a wider scattered RGB, which may indicate a substantial age and/or 
metallicity spread.
One can clearly see the almost complete absence of upper main sequence stars
in the CMDs indicating a lack of ongoing star formation in the dSph galaxies. 
A denser concentration of red clump (RC) stars of a slightly elongated 
shape is conspicuous in the range of $F814W$ ~ 24--26 mag (26--27 mag for the more
distant KK\,258 and KKs\,03). With the exception of the shallower CMDs of the 
isolated dSphs KKR\,25, KK\,258 and KKs\,03, the horizontal branch (HB) extending
to the left from RC are also visible at the diagrams. The HBs population density 
of these branches varies significantly for the different sample galaxies.

\section{Star formation history determination}

\subsection{The method}
Star formation history of the selected dwarf spheroidal galaxies was 
determined using our StarProbe \citep{mm2004} software package. In the works 
published for a part of the galaxy sample (see, for example, \citet{mak2017}), 
our technics was described in sufficient detail. Here we briefly describe the basic 
principles and steps of our calculations.

Observed distribution of stars in a colour-magnitude diagram is a linear 
superposition of all stars at the various evolutionary stages, which were formed 
in the galaxy during its life. A number of additional parameters also have 
a strong influence on this photometric distribution: distance to the object, 
external and internal extinction, and photometric errors. Accurate photometric 
distances were estimated for all objects of our sample (see Sect.~\ref{sec:sample}). 
We also performed highly accurate estimates of photometric errors for all galaxies 
using extensive tests with artificial stars (see Sect.~\ref{phot}). The Galactic
extinction values were taken from \citet{schlafly}. The internal extinction
in dwarf spheroidal galaxies is usually negligible due to lack of gas and regions
of active star formation.

Observational data and model data are used in the form of a Hess diagram, which
is a two-dimensional histogram representing the number of stars in a certain
range (bin) of a magnitude and a colour. Model Hess diagrams are built from
theoretical stellar isochrones. Each isochrone corresponds to a specific age
and metallicity, and the entire set of these models fits a fairly wide 
range of ages and metallicities of the stellar populations. The SFH determination
comes down to finding of a linear combination of partial model CMDs that approximate
the observational data best. We construct the analytical distribution function of
stars at the Hess diagram for each isochrone, taking into account initial mass
function (IMF), photometric errors, bin size of the Hess diagram, distance modulus
and extinction. We use the Padova stellar isochrone calculations made with PARSEC 
library\footnote{\url{http://stev.oapd.inaf.it/cgi-bin/cmd}} and the Salpeter IMF:
$\rho(m)\,dm\sim m^{-2.35}dm$. The chosen normalization gives the total integral 
probability equal to one in the range of 0.1--100~$M_\odot$. For a given isochrone, 
the mass of a star completely determines its observational characteristics and, 
together with the parameters mentioned above fully determines the probability of a star
of the given mass to occur in any bin of the Hess diagram.
Individual isochrones of the same metallicity for a given period of time
are combined together assuming a constant star formation rate during this period.

To determine SFH, we firstly find the most significant variables (i.e., partial
model CMDs) that differ from zero with a given probability.
Then we determine values of the significant variables by the maximum likelihood method.
Thus, we construct the maximum likelihood function to get the desired solution using 
well-known minimization algorithms.

\subsection{SFH measurement results}

\begin{figure*}
\begin{tabular}{cc}
\includegraphics[height=5cm]{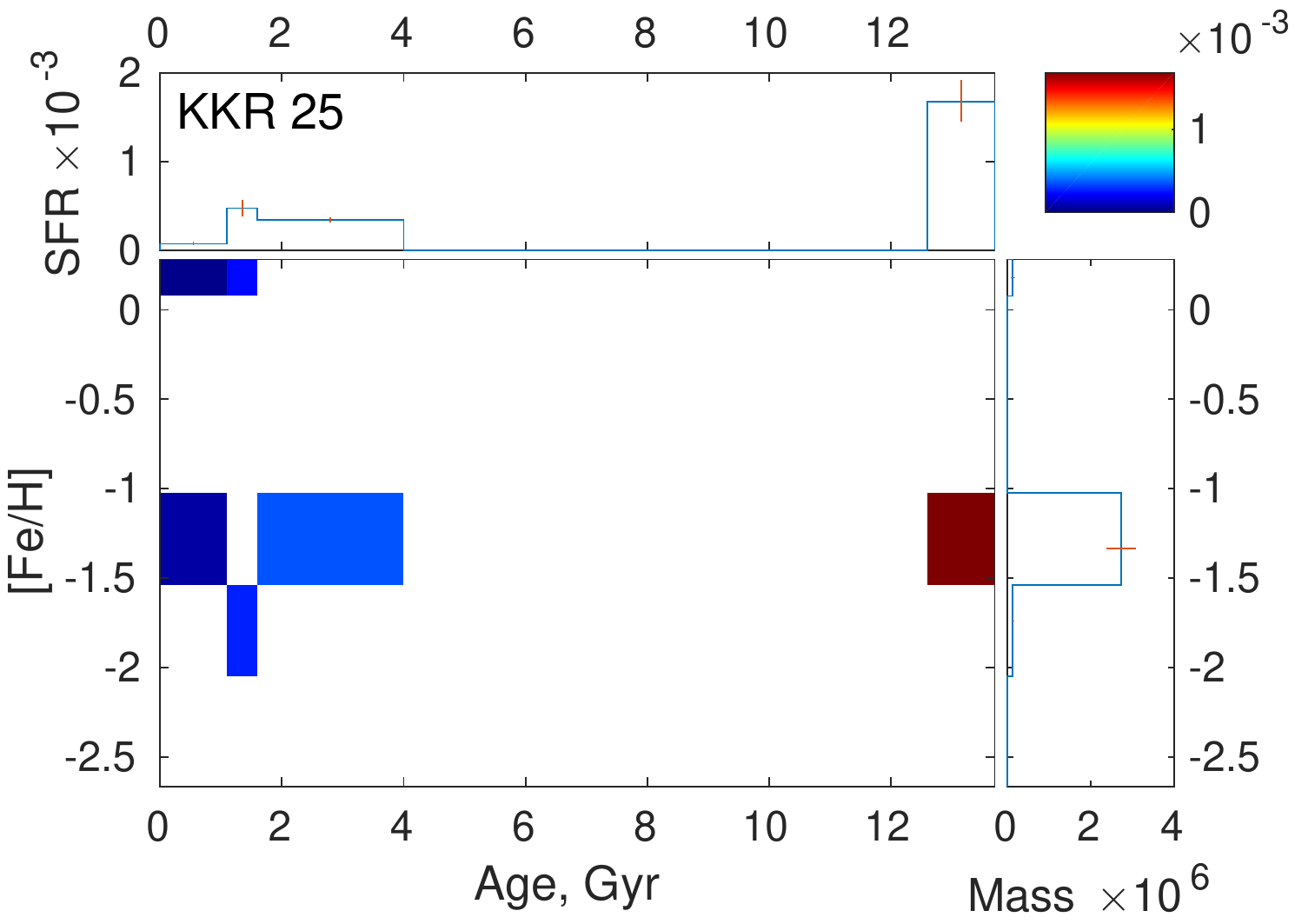} &
\includegraphics[height=5cm]{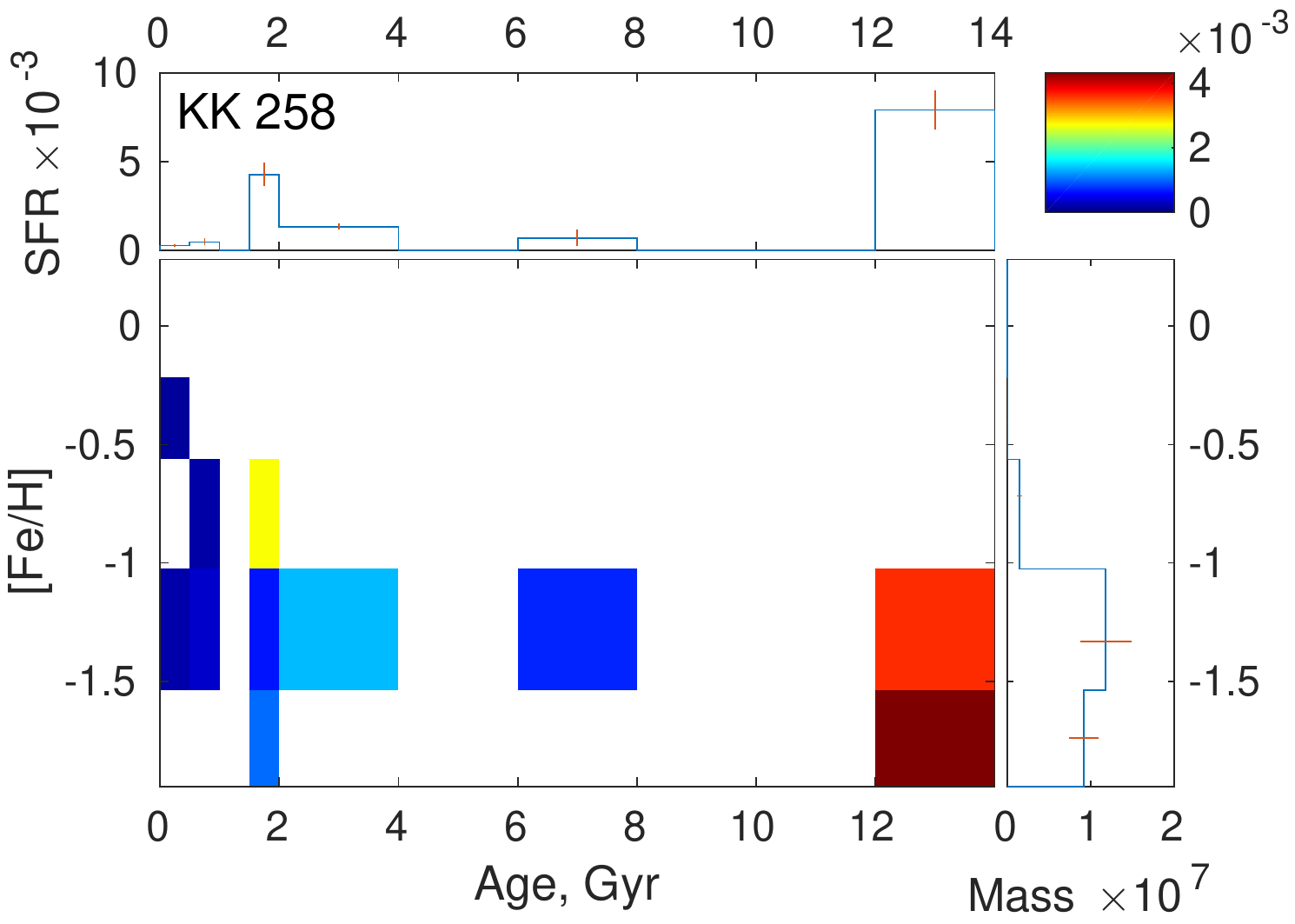}\\
\includegraphics[height=5cm]{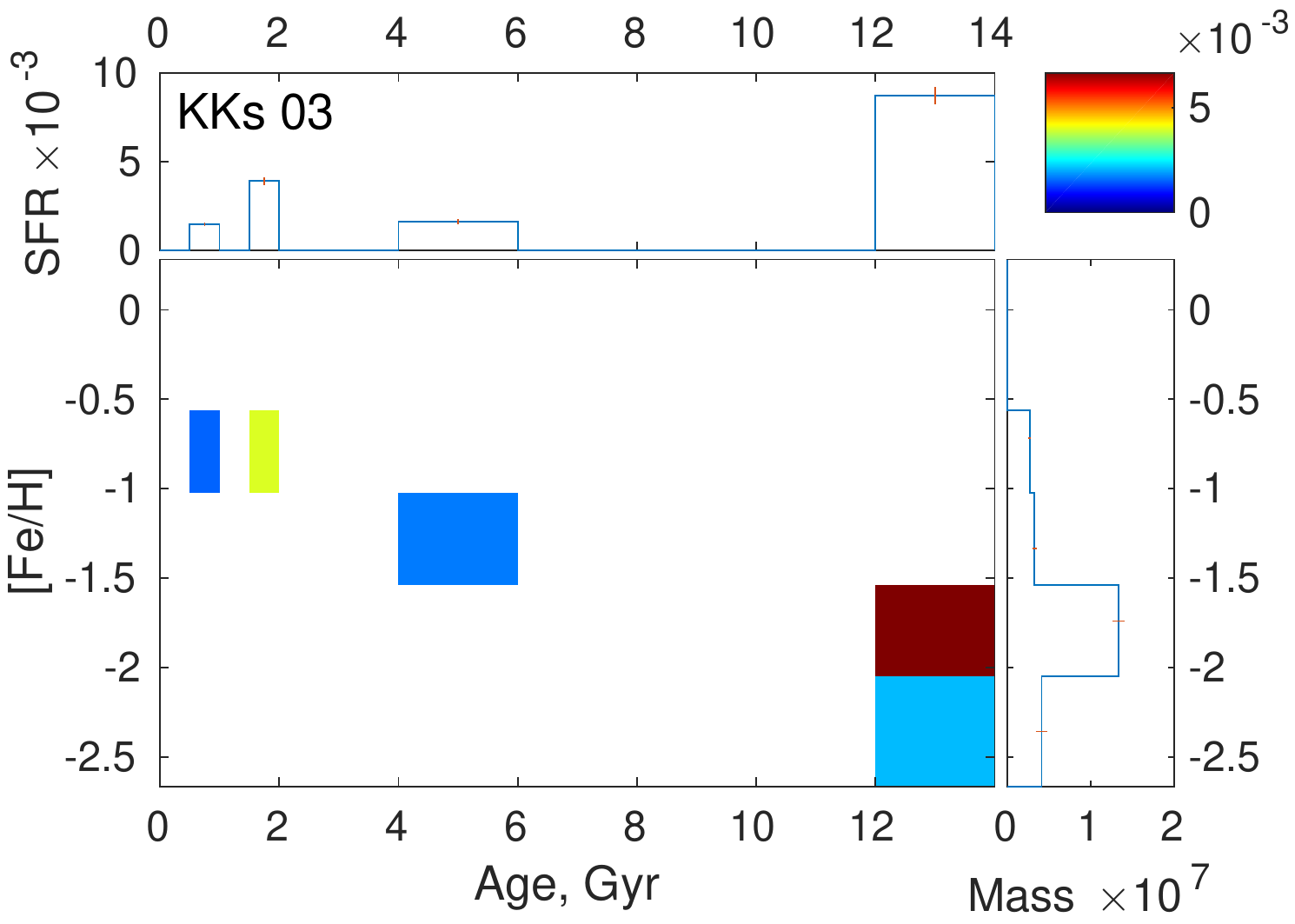} &
\includegraphics[height=5cm]{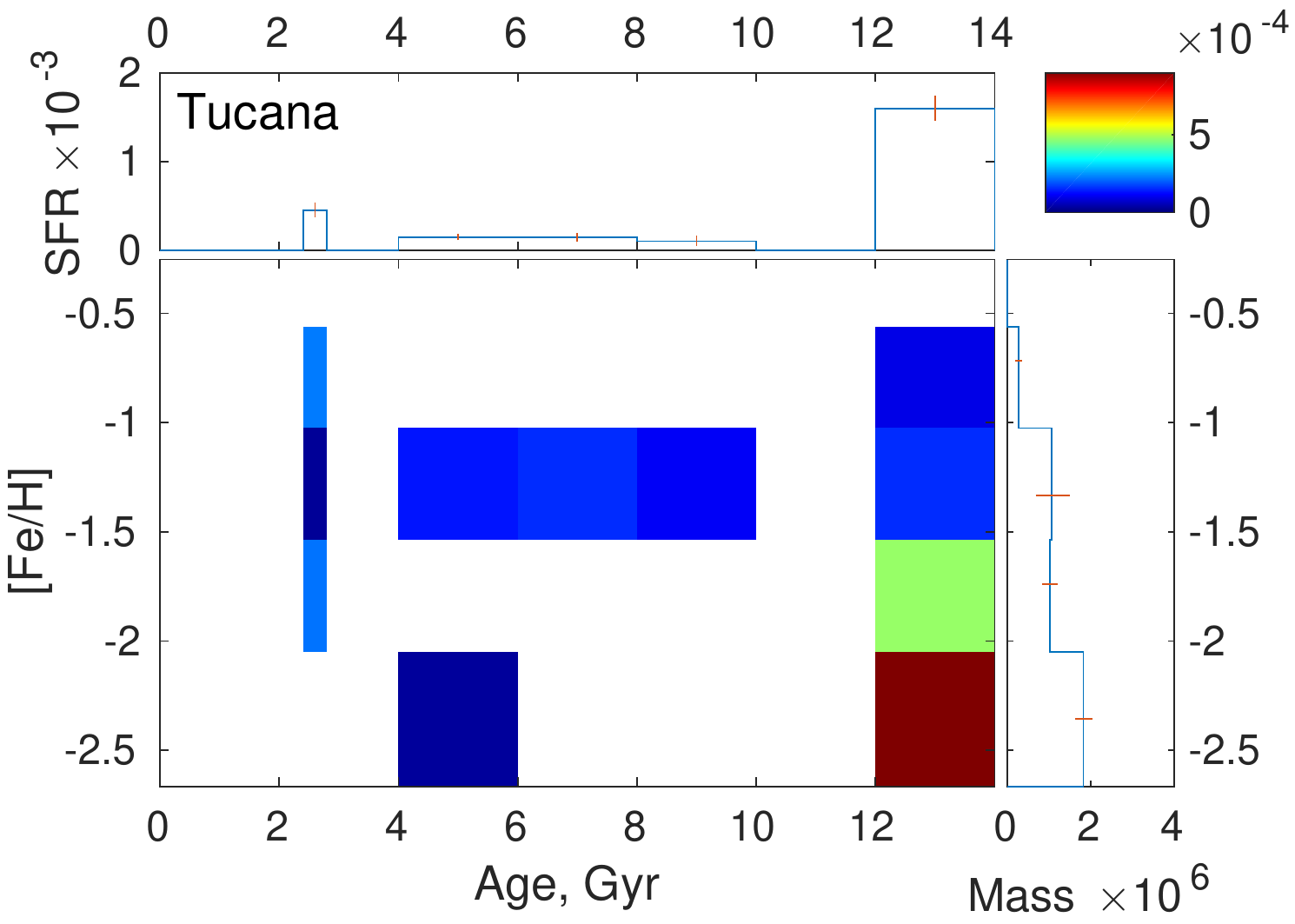} \\
\includegraphics[height=5cm]{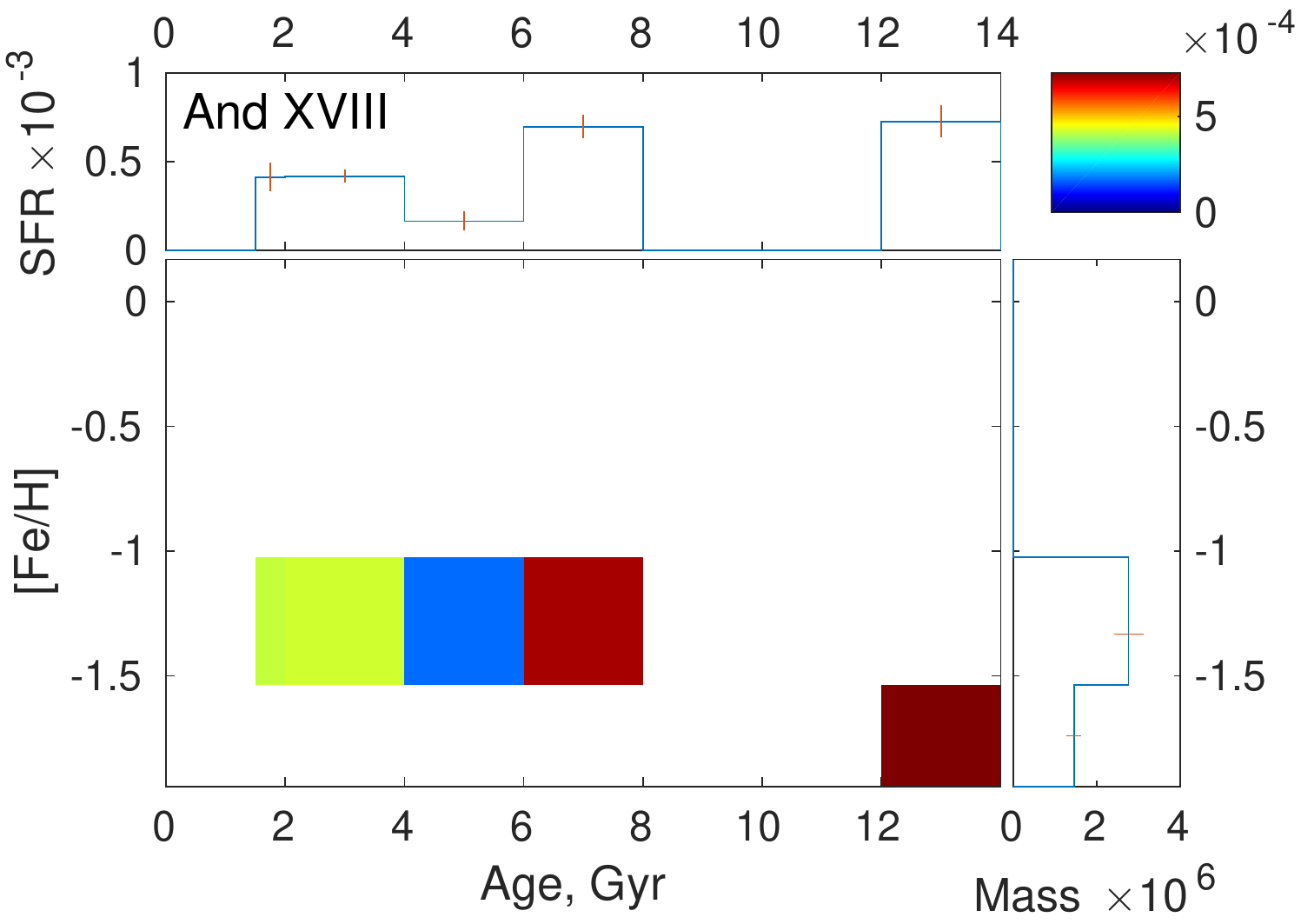} & 
\includegraphics[height=5cm]{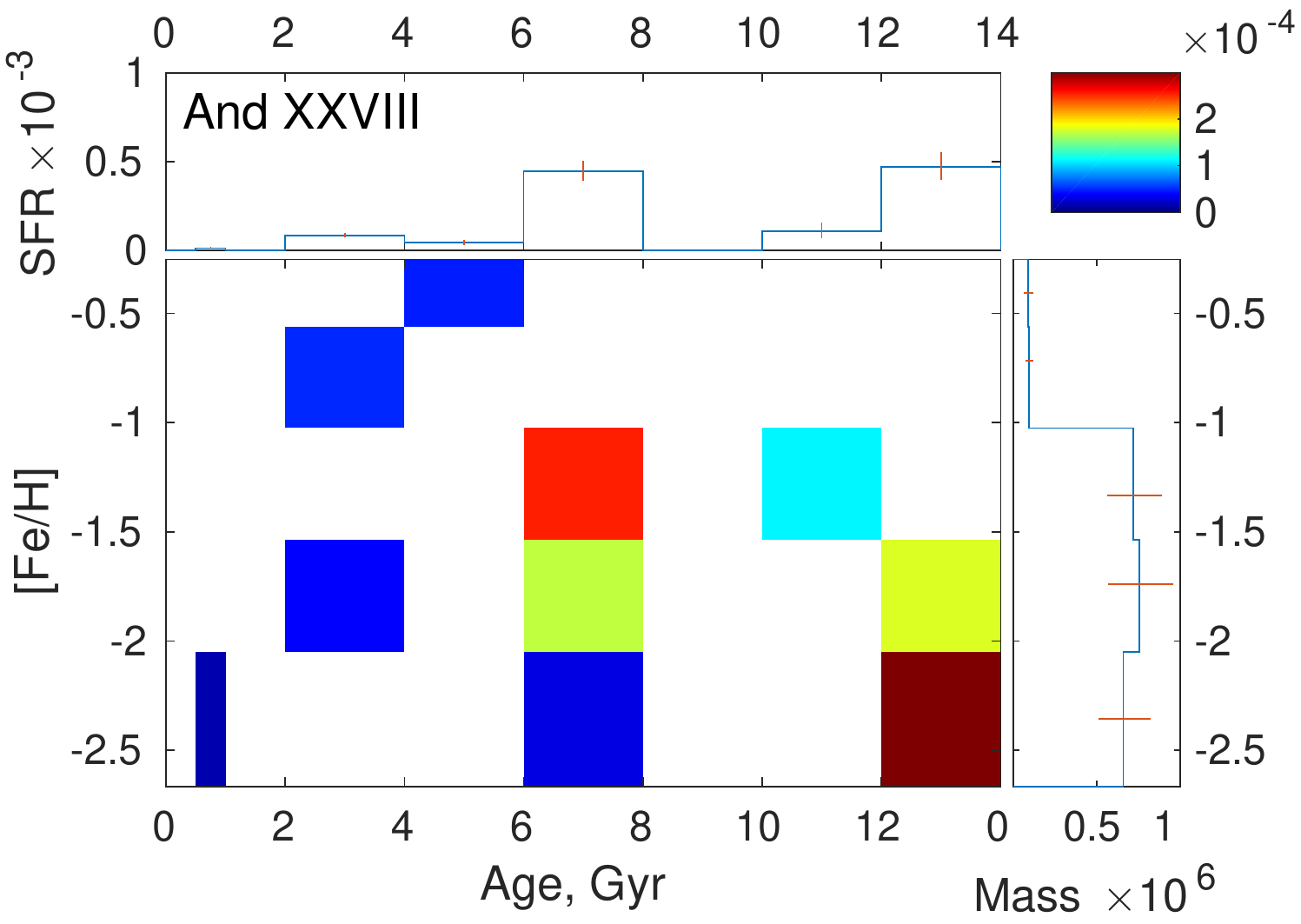} \\
\includegraphics[height=5cm]{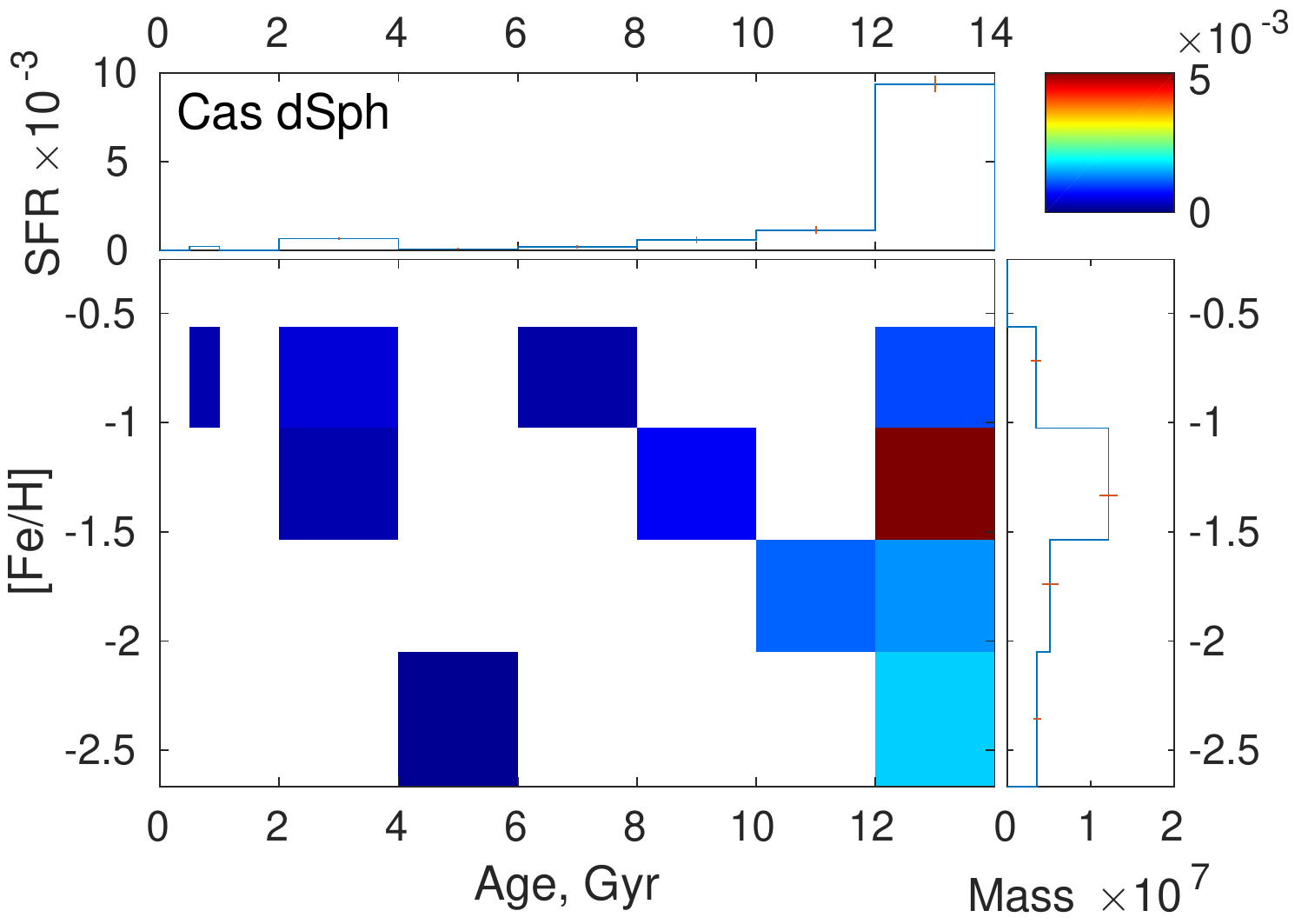} &
\includegraphics[height=5cm]{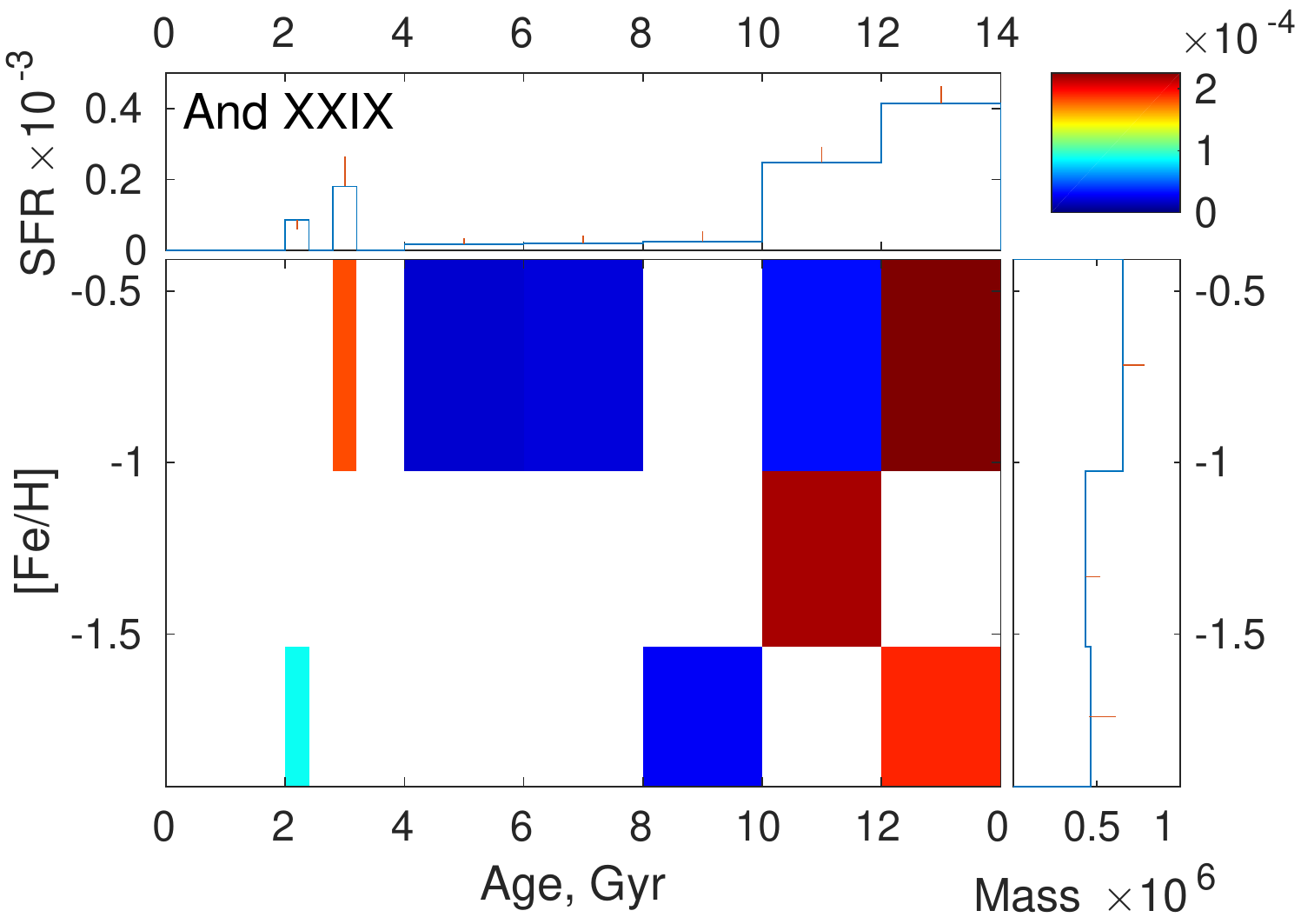} \\
\end{tabular}
\caption{Star formation histories of the dSph galaxies of our sample. Three
panels are given for each galaxy: the upper one shows the star formation rate
(SFR) dependence (in the units of solar masses per year, $M_{\odot}$\,$yr^{-1}$) on
the age of the stellar populations (in billions of years, Gyr); the bottom 
panel represents the metallicity [Fe/H] of the stellar component as a function
of age, where the coloured rectangles correspond to the estimated periods
of star formation; the right panel shows the measured mass of stars of the 
respective metallicity. The formal errors of our calculations are indicated
as vertical bars.
}
\label{fig:sfh}
\end{figure*}

The results of the SFH reconstruction for our sample 
of galaxies are presented in Fig.~\ref{fig:sfh} and in Table~\ref{tab:sfmes}. 
The columns in the Table~\ref{tab:sfmes} are: (1) -- galaxy name; (2) -- metallicity of the SF episode; 
(3) -- mean age of the current SF episode;  
(4) -- mean SFR within the current SF episode, in square brackets there are lower and higher 
$1\sigma$ errors of the estimated SFR derived from the computed confidence levels; 
(4) -- current SF episode duration; (5) -- total stellar mass of the stars formed in 
the current SF episode. We have estimated the star 
formation rate (SFR) and metallicity depending on the age of the stellar 
populations, as well as the stellar mass formed in each related period of
star formation. A number of common trends follows from these results:
the star formation has been quenched over the last 1--2 Gyrs which is quite consistent
with the morphological type (dSph and dTr) of the studied dwarfs; there is 
an initial active burst of star formation from about 
10--12 to 13.8 Gyrs ago. The fraction of stars formed during this 
period is different for our objects (we will discuss it below). In addition, 
there are stars formed about 4--8 Gyrs ago. The periods and intensity of star 
formation at these `middle' ages can vary significantly from object to object. 
The metallicity of stars is mostly low, and the metal enrichment during the
galaxy life looks insignificant, quite typically for faint dwarf spheroidals
(as well as for dwarf irregulars) (e.g. \citet{grebel2001, tolstoy2009}).

\begin{table*}
\centering
\caption{Star formation parameters of the studied galaxies.}
\label{tab:sfmes}
\scriptsize
\begin{tabular}{lccccc}
Name       & Fe/H   & T     &   SFR                          &   $\Delta T$  & M  \\ 
           &        & Gyr   &  $M_{\odot}$\,$yr^{-1}$        &   Gyr & $M_{\odot}$ \\
\hline
KKR 25     & --1.74 & 1.35  & 0.000256 [0.000199-- 0.000317] &  0.50 & 1.28e+05  \\
           & --1.33 & 0.55  & 5.72e-05 [4.18e-05-- 7.48e-05] &  1.10 & 6.29e+04  \\
           & --1.33 & 2.80  & 0.000342 [0.000312-- 0.000372] &  2.40 & 8.20e+05  \\
           & --1.33 & 13.15 & 0.00168  [0.00145 -- 0.00192 ] &  1.10 & 1.85e+06  \\
           &   0.18 & 0.55  & 1.74e-05 [7.25e-06-- 3.14e-05] &  1.10 & 1.92e+04  \\
           &   0.18 & 1.35  & 0.000222 [0.000153-- 0.000297] &  0.50 & 1.11e+05  \\
\hline
KK 258     & --1.74 & 1.75 & 0.000986 [0.000701-- 0.00128 ]  & 0.50 & 4.93e+05  \\
           & --1.74 &13.00 & 0.00433  [0.00355 -- 0.00513 ]  & 2.00 & 8.66e+06  \\
           & --1.33 & 0.25 & 0.000176 [0.000115-- 0.000247]  & 0.50 & 8.78e+04  \\
           & --1.33 & 0.75 & 0.000308 [0.000202-- 0.000425]  & 0.50 & 1.54e+05  \\
           & --1.33 & 1.75 & 0.000612 [0.000273-- 0.000968]  & 0.50 & 3.06e+05  \\
           & --1.33 & 3.00 & 0.00133  [0.00116 -- 0.00151 ]  & 2.00 & 2.67e+06  \\
           & --1.33 & 7.00 & 0.000693 [0.000231-- 0.00117 ]  & 2.00 & 1.39e+06  \\
           & --1.33 &13.00 & 0.00358  [0.00283 -- 0.00438 ]  & 2.00 & 7.17e+06  \\
           & --0.72 & 0.75 & 0.000169 [1.87e-05-- 0.00038 ]  & 0.50 & 8.44e+04  \\
           & --0.72 & 1.75 & 0.00268  [0.00223 -- 0.00316 ]  & 0.50 & 1.34e+06  \\
           & --0.41 & 0.25 & 0.000107 [6.62e-05-- 0.000155]  & 0.50 & 5.33e+04  \\
\hline
KKs 03     & --2.36 &13.00 & 0.00206  [0.00172 -- 0.00241 ]  & 2.00 & 4.12e+06  \\
           & --1.74 &13.00 & 0.00667  [0.00632 -- 0.00702 ]  & 2.00 & 1.33e+07  \\
           & --1.33 & 5.00 & 0.00163  [0.00149 -- 0.00176 ]  & 2.00 & 3.25e+06  \\
           & --0.72 & 0.75 & 0.00147  [0.00136 -- 0.00159 ]  & 0.50 & 7.37e+05  \\
           & --0.72 & 1.75 & 0.00391  [0.0037  -- 0.00414 ]  & 0.50 & 1.96e+06  \\
\hline
Tucana    & --2.36 & 5.00 & 2.1e-05  [7.04e-06-- 3.91e-05] &  2.00 & 4.20e+04  \\
          & --2.36 &13.00 & 0.000893 [0.000808-- 0.000982] &  2.00 & 1.79e+06  \\
          & --1.74 & 2.60 & 0.00021  [0.000171-- 0.000255] &  0.40 & 8.42e+04  \\
          & --1.74 &13.00 & 0.000466 [0.000385-- 0.000552] &  2.00 & 9.32e+05  \\
          & --1.33 & 2.60 & 2.09e-05 [5.29e-08-- 6.75e-05] &  0.40 & 8.34e+03  \\
          & --1.33 & 5.00 & 0.000129 [9.76e-05-- 0.000161] &  2.00 & 2.57e+05  \\
          & --1.33 & 7.00 & 0.000148 [9.73e-05-- 0.000203] &  2.00 & 2.97e+05  \\
          & --1.33 & 9.00 & 0.000104 [4.91e-05-- 0.000165] &  2.00 & 2.09e+05  \\
          & --1.33 &13.00 & 0.000149 [9.51e-05-- 0.000211] &  2.00 & 2.98e+05  \\
          & --0.72 & 2.60 & 0.000218 [0.000161-- 0.000281] &  0.40 & 8.74e+04  \\
          & --0.72 &13.00 & 8.96e-05 [6.02e-05-- 0.000125] &  2.00 & 1.79e+05  \\
\hline
And XVIII & --1.74 &13.00 & 0.000728 [0.00064 -- 0.000819] &  2.00 & 1.46e+06  \\
          & --1.33 & 1.75 & 0.000411 [0.000332-- 0.000495] &  0.50 & 2.06e+05  \\
          & --1.33 & 3.00 & 0.000418 [0.000381-- 0.000456] &  2.00 & 8.37e+05  \\
          & --1.33 & 5.00 & 0.000165 [0.000111-- 0.000221] &  2.00 & 3.31e+05  \\
          & --1.33 & 7.00 & 0.000697 [0.000631-- 0.000765] &  2.00 & 1.39e+06  \\
\hline
And XXVIII & --2.36 & 0.75 & 1.29e-05 [8.44e-06-- 1.85e-05] &  0.50 & 6.46e+03  \\
           & --2.36 & 7.00 & 2.9e-05  [8.02e-06-- 5.37e-05] &  2.00 & 5.80e+04  \\
           & --2.36 &13.00 & 0.000298 [0.000245-- 0.000354] &  2.00 & 5.95e+05  \\
           & --1.74 & 3.00 & 3.67e-05 [2.96e-05-- 4.45e-05] &  2.00 & 7.35e+04  \\
           & --1.74 & 7.00 & 0.000167 [0.000133-- 0.000202] &  2.00 & 3.34e+05  \\
           & --1.74 &13.00 & 0.000175 [0.000122-- 0.000234] &  2.00 & 3.50e+05  \\
           & --1.33 & 7.00 & 0.00025  [0.000212-- 0.000289] &  2.00 & 5.00e+05  \\
           & --1.33 &11.00 & 0.000109 [6.82e-05-- 0.000156] &  2.00 & 2.18e+05  \\
           & --0.72 & 3.00 & 4.8e-05  [3.78e-05-- 5.88e-05] &  2.00 & 9.60e+04  \\
           & --0.41 & 5.00 & 4.43e-05 [3.13e-05-- 5.88e-05] &  2.00 & 8.85e+04  \\
\hline
Cas dSph   & --2.36 & 5.00 & 8.38e-05 [4.12e-05-- 0.000133] &  2.00 & 1.68e+05  \\
           & --2.36 &13.00 & 0.0017   [0.0015  -- 0.00191 ] &  2.00 & 3.41e+06  \\
           & --1.74 &11.00 & 0.00115  [0.000917-- 0.00139 ] &  2.00 & 2.30e+06  \\
           & --1.74 &13.00 & 0.00141  [0.00117 -- 0.00167 ] &  2.00 & 2.82e+06  \\
           & --1.33 & 3.00 & 0.000233 [0.000194-- 0.000273] &  2.00 & 4.65e+05  \\
           & --1.33 & 9.00 & 0.000605 [0.000415-- 0.000807] &  2.00 & 1.21e+06  \\
           & --1.33 &13.00 & 0.00524  [0.00494 -- 0.00555 ] &  2.00 & 1.05e+07  \\
           & --0.72 & 0.75 & 0.000233 [0.000209-- 0.000258] &  0.50 & 1.16e+05  \\
           & --0.72 & 3.00 & 0.000442 [0.000389-- 0.000499] &  2.00 & 8.85e+05  \\
           & --0.72 & 7.00 & 0.0002   [0.000115-- 0.000297] &  2.00 & 4.00e+05  \\
           & --0.72 &13.00 & 0.00101  [0.000861-- 0.00117 ] &  2.00 & 2.02e+06  \\
\hline
And XXIX   & --1.74 & 2.20 & 8.7e-05  [6.04e-05-- 8.7e-05 ] &  0.40 & 3.48e+04  \\
           & --1.74 & 9.00 & 2.57e-05 [2.57e-05-- 5.54e-05] &  2.00 & 5.13e+04  \\
           & --1.74 &13.00 & 0.000189 [0.000189-- 0.000236] &  2.00 & 3.78e+05  \\
           & --1.33 &11.00 & 0.000216 [0.000216-- 0.00026 ] &  2.00 & 4.33e+05  \\
           & --0.72 & 3.00 & 0.000181 [0.000181-- 0.000266] &  0.40 & 7.22e+04  \\
           & --0.72 & 5.00 & 1.71e-05 [1.71e-05-- 3.53e-05] &  2.00 & 3.42e+04  \\
           & --0.72 & 7.00 & 1.97e-05 [1.97e-05-- 4.25e-05] &  2.00 & 3.94e+04  \\
           & --0.72 &11.00 & 3.09e-05 [3.09e-05-- 3.49e-05] &  2.00 & 6.17e+04  \\
           & --0.72 &13.00 & 0.000226 [0.000226-- 0.000228] &  2.00 & 4.52e+05  \\
\hline
\end{tabular}
\end{table*}

\section{Comparative analysis of the results}

We performed comparative analysis of the obtained SFHs 
of the sample galaxies located at different distances from 
the Andromeda galaxy.
The results 
of this analysis are presented in Table~\ref{tab:res}. The first column of 
the table gives a galaxy name; (2) is the distance from M\,31; (3) -- the mass
of stars formed in the last 2 Gyr, given as a percentage of the total stellar 
mass; (4) -- the mass of stars older than 8 Gyr; (5) -- the measured total stellar
mass of the galaxy; (6) - SFR in the first 12--13.7 Gyr.

\begin{table*}
\centering
\caption{Star formation parameters of the studied galaxies.}
\label{tab:res}
\begin{tabular}{ccllcc}
Name         & $D_{M31}$     & $M^*_{\leq2}$ & $M^*_{\geq8}$ & $M^*_{Total}$ & $SFR_{\geq12}$  \\ 
            &  Mpc          &    per cent            &    per cent             &   $M_{\odot}$ & $M_{\odot}$\,$yr^{-1}$  \\
\hline
KKR 25      & 1.93$\pm$0.07 & 11   & 62 & $3.0\cdot10^6$ & $1.7\pm0.2\cdot10^{-3}$ \\
KK 258      & 0.84$\pm$0.09 & 11   & 70 & $2.2\cdot10^7$ & $7.9\pm4.0\cdot10^{-3}$ \\
KKs 03      & 2.12$\pm$0.07 & 12   & 74 & $2.3\cdot10^7$ & $8.7\pm0.4\cdot10^{-3}$ \\
\multicolumn{6}{c}{\dotfill} \\
Tucana      & 0.92$\pm$0.02 &  0   & 81 & $4.2\cdot10^6$ & $1.6\pm0.2\cdot10^{-3}$ \\
And XVIII   & 0.58$\pm$0.09 &  5   & 34 & $4.2\cdot10^6$ & $7.3\pm0.9\cdot10^{-4}$ \\
\multicolumn{6}{c}{\dotfill} \\
And XXVIII  & 0.38$\pm$0.09 &  0   & 50 & $2.3\cdot10^6$ & $4.7\pm1.2\cdot10^{-4}$ \\
Cas dSph    & 0.23$\pm$0.03 & 0.5  & 93 & $2.4\cdot10^7$ & $9.4\pm0.8\cdot10^{-3}$ \\
And XXIX    & 0.20$\pm$0.02 &  0   & 86 & $1.6\cdot10^6$ & $4.2\pm1.3\cdot10^{-4}$ \\
\hline
\end{tabular}
\end{table*}

\begin{figure*}
\centering
\includegraphics[height=6cm]{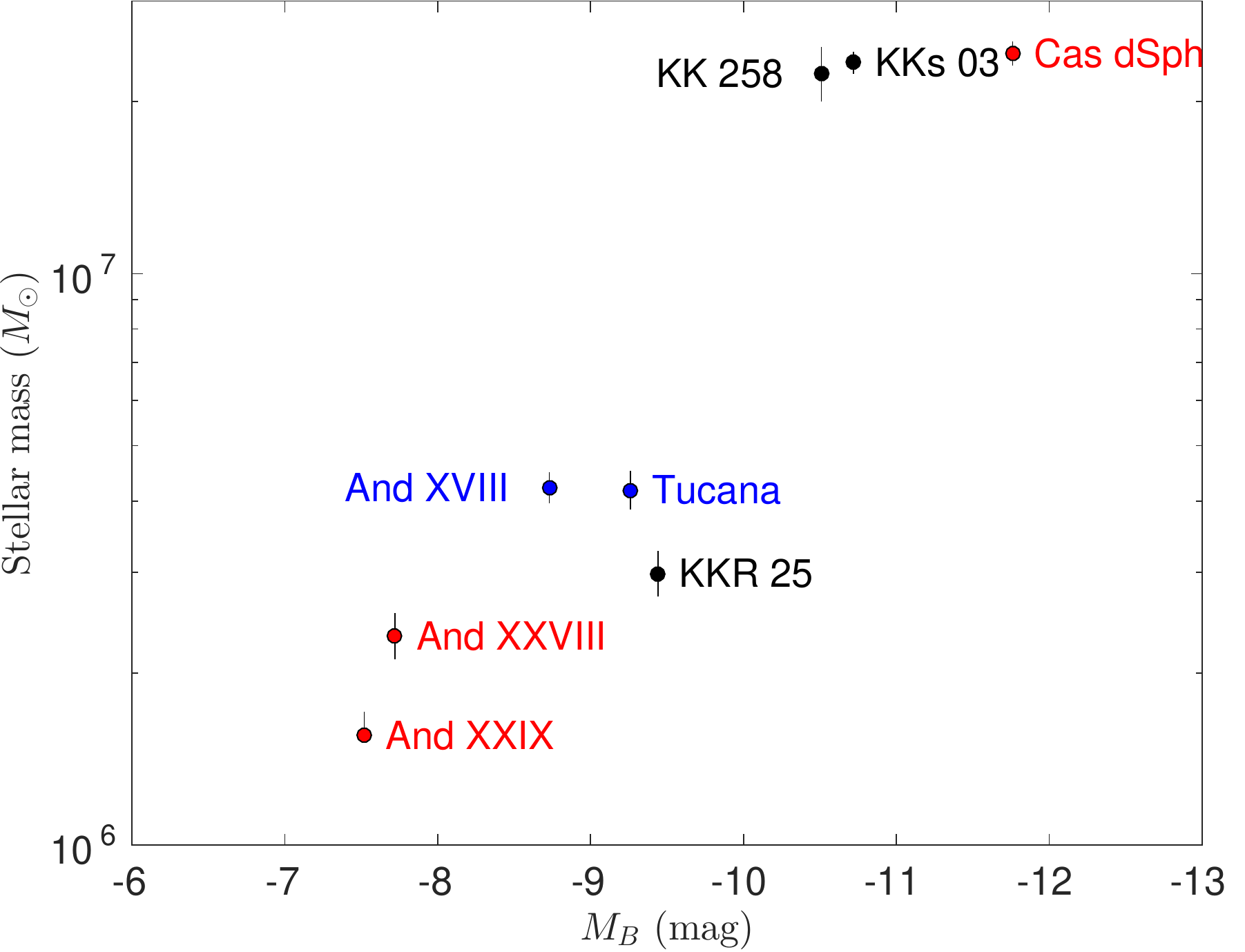}
\includegraphics[height=6cm]{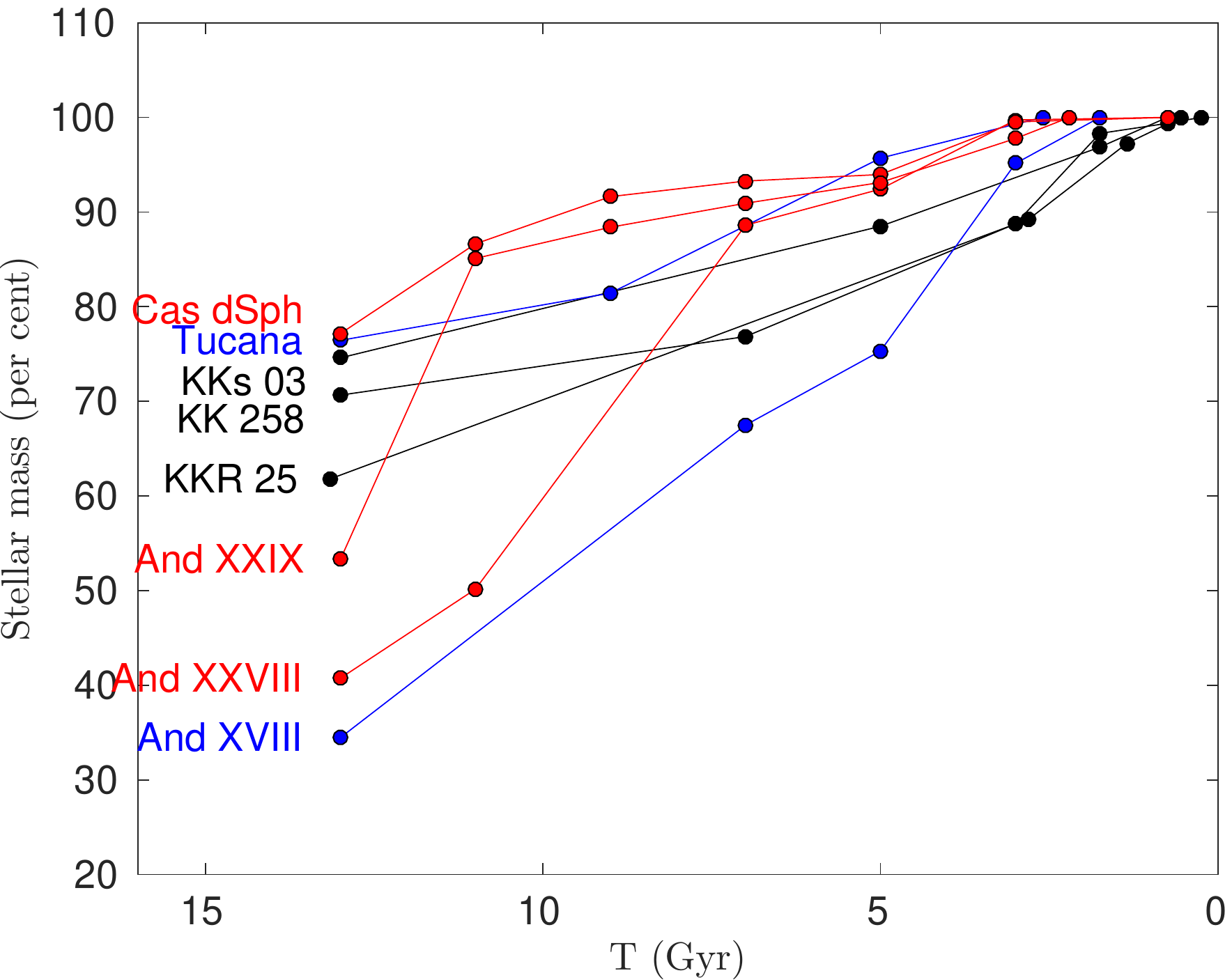}
\includegraphics[height=6cm]{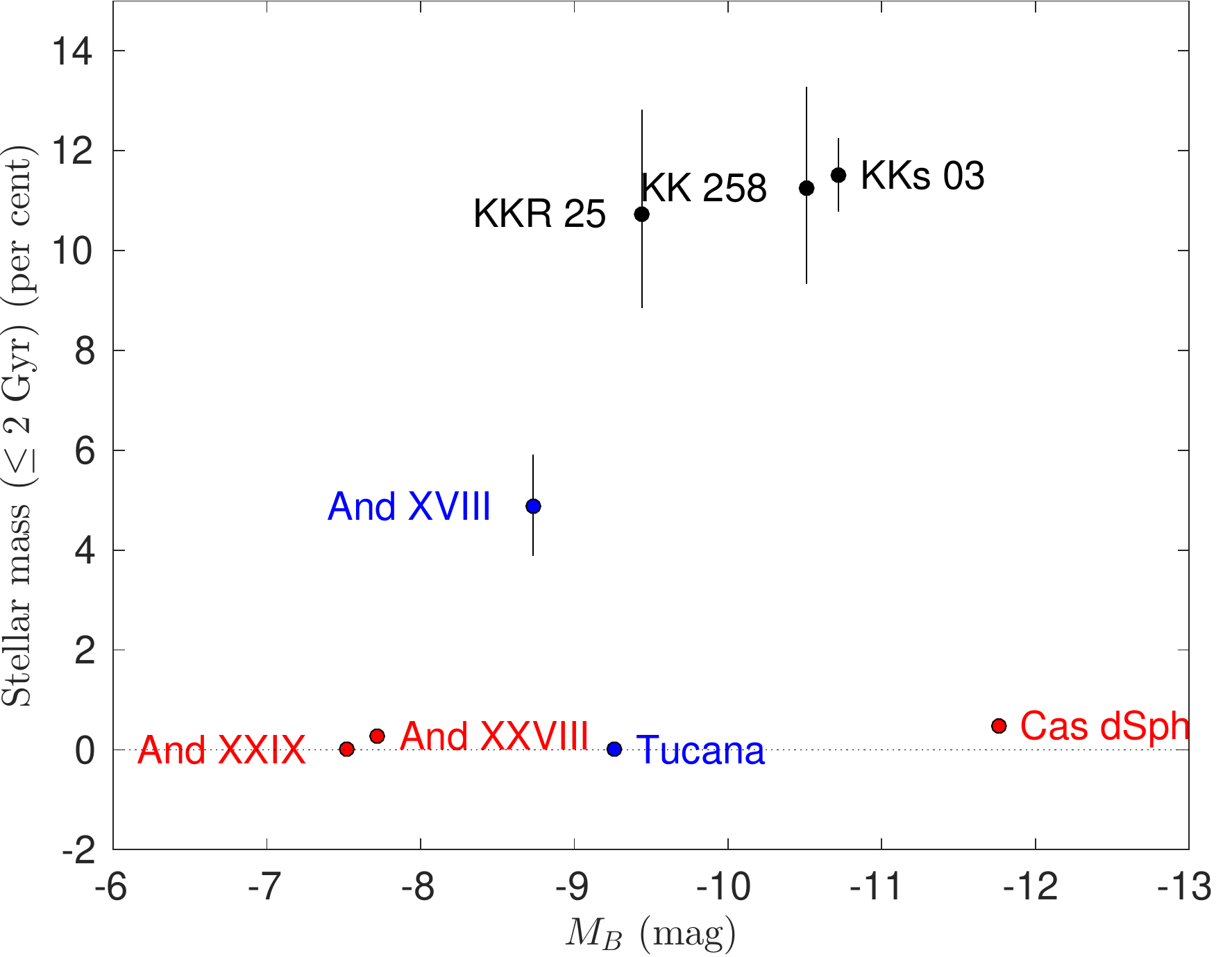}
\caption{The upper left panel shows the measured total stellar
mass for a sample of dwarf galaxies vs. the absolute magnitude in filter B;
the upper right panel represents the cumulative stellar mass function for our sample;
the lower panel of shows the stellar mass formed in the last
2 Gyr vs. the galaxy absolute magnitudes. Every galaxy
signed, and in the colour version of this figure the isolated dwarf
spheroidals are indicated in black, within the zero velocity radius of the
M\,31 -- blue, and within the virial radius -- red.
}
\label{fig:res}
\end{figure*}

Some relations reflecting star formation activity of our objects are presented
in Fig.~\ref{fig:res}. The upper left panel shows estimated total stellar mass 
with respect to the absolute stellar magnitude B in filter. The stellar mass range
of the studied galaxies is more than the order of magnitude from about $10^6$ 
to $2.5\cdot 10^7$, quite typical for classic nearby dwarfs. Higher stellar masses naturally 
correspond to brighter galaxies. 

The upper right panel presents the cumulative mass 
function (growth of the formed stellar mass with time) as a percentage
of the total stellar mass of the galaxy. 
Note that a dwarf galaxy of higher luminosity 
forms the larger part of the stars ($\geq$60 per cent) during the initial outburst 
of star formation ($\geq 12$ Gyr ago). An interesting exception is the Tucana dwarf.
Following the relation found above, the stellar mass of this galaxy should be 
higher. Nevertheless, it is known that in the past the object could experience close 
approach to the central body of the group, and possibly interactions with other
dwarfs before its fall on the Local Group. All these processes could cause
the loss of stellar mass and material for star formation \citep{monelli2010}.
From the other side, the Holmberg diameter of Tucana can be underestimated \citep{saviane1996},
and part of the galaxy falls outside the ACS field. In this case the total stellar mass
of Tucana dSph should be underestimated.
One of the recent detailed SFH determination of the Tucana dSph was performed by
\citet{savino2019} using HST/ACS photometry. The colour-magnitude diagram in
this study is deeper than ours, and also the authors put finer scale for their Age--[Fe/H] plane.
Nevertheless, the overall star formation periods and the metallicity range
of the SFH by \citet{savino2019} and our measurements are consistent. The initial burst
of star formation occurred about 11--14 Gyr ago (12--14 Gyr in the present work),
small number of stars continued to form about 6--11 Gyr ago (4--10 Gyr ago) and the
residual recent star formation appears 3--4 Gyr ago (2.5--3 Gyr ago). Star formation rate of
this last period (roughly estimated from the Fig.\,8 of \citet{savino2019}) is about
$0.5-0.6\cdot 10^{-4}$ $M_{\odot}$\,$yr^{-1}$, in good agreement with the SFR of about $0.4\cdot 10^{-4}$
$M_{\odot}$\,$yr^{-1}$ measured in our work. The both measurements show the comparable
total stellar mass $3.13\cdot 10^6$ $M_{\odot}$ ($4.2\cdot 10^6$ $M_{\odot}$).

The bottom panel of Fig.~\ref{fig:res} shows the stellar mass formed in the last
2 Gyr vs. the galaxy absolute magnitudes. The percentage of the mass formed 
is small for all the objects. This plot could demonstrate how intensive
is the residual star formation in dwarf spheroidal galaxies depending on the degree
of isolation of the object. Highly isolated galaxies clearly show recent residual
star formation ($>10$ per cent of the total stellar mass), while within the virial radius
it is practically zero. And\,XVIII and Tucana are situated beyond the virial 
radius (but within the Local Group zero velocity sphere). Following 
the logic of the above plot, they should be located between those two extreme
cases. Nonetheless, Tucana is again out of the trend, located among the galaxies
inside the virial radius. It also could be an argument in favour that in the past
this dwarf was among the closer satellites of M\,31. 

\citet{simpson2017} predicted quenching of star formation in satellite galaxies 
of the Local Group from a suite of 30 cosmological zoom simulations of MW-like 
host galaxies. The authors found that systems within host galaxy $R_{200}$ 
are comprised of two populations: first infall and `backsplash' galaxies 
that have had a much more extended interaction with the host. Backsplash galaxies 
that do not returned to the host by z=0, exhibit quenching properties similar 
to galaxies within $R_{200}$ and are distinct from other external systems.
The last scenario are well match with the star formation properties of Tucana dSph.

It should be noted that the results are affected by the well known problem of
age-metallicity degeneration in the process of SFH measurement, and some 
dependence of the results on the isochrone set used \citep{wil2017}. 
Besides, the obtained colour-magnitude diagrams are somewhat not deep enough,
in the sense that the photometric limit is brighter than MSTO (Main Sequence
turn off). At the same time, uniform data set and its homogeneous processing
should reduce the influence of these factors on the results of our comparative
analysis to a minimum.

When analysing SFHs of dwarf satellite galaxies in the nearby galaxy
groups (first of all, the Local Group), one should take into account,
that even in a particular mass range and fixed environment, there is a large 
degree of scatter in the SFHs, frequently noted in various researches
(see, for example, \citet{garrison2019}). The mentioned study has used
a set of about 500 dwarf galaxies taken from FIRE-2 simulations to explore
how their SFHs vary with environment. They conclude, that the satellites
(d$\leq$300 kpc) tend to form their stars earlier than equivalent
$300 < d < 1000$~kpc dwarfs, consistent with the picture, where interactions
with MW-mass hosts inhibit star formation. 
A galaxy could loss its gas due to ram pressure effects and tidal interaction
effects within the virial radius of the group (Andromeda galaxy group). This
scenario is well established and commonly used for the explanation of the star
formation quenching of dwarf spheroidal galaxies (see the paper of \citet{mak2017}
mentioned above and the references therein).

\citet{garrison2019} also found, that highly isolated dwarf galaxies
with $M_*\leq 10^7 M_{\odot}$ quenched even later, than dwarfs with
$300 < d < 1000$~kpc dwarfs. These conclusions are consistent with the picture
demonstrated by the dwarf galaxies of our sample.
At the same time, \citet{simpson2017} found strong mass-dependent and 
distance-dependent quenching, where dwarf systems beyond 600 kpc are 
only strongly quenched below a stellar mass of $10^7$ $M_{\odot}$. In general, our
sample dwarfs are following this trend, with the apparent exception of KKR\,25.
However, this dwarf spheroidal galaxy is extremely isolated, and probably
other mechanisms of star formation quenching should work there. 

\section{Conclusions}

In our study we performed a comparative analysis of highly homogeneous observational
and photometric data of 8 dwarf spheroidal galaxies of the Local Group (M\,31 galaxy subgroup).
We provided uniform processing and stellar photometry of images, as well as
homogeneous star formation history measurements, to study the dwarf spheroidal galaxy 
evolution and possible star formation quenching mechanisms of satellites
situated at the different distances from the host galaxy.

We estimated the star formation rate and metallicity depending on the age of the stellar 
populations, as well as the stellar mass formed in each related period of
star formation. We found that the star formation has been quenched over the 
last 1--2 Gyr. All the studied objects show an initial active burst of star formation 
from about 10--12 to 13.8 Gyr ago. In addition, 
there are stars formed about 4--8 Gyrs ago. The periods and intensity of star 
formation at these `middle' ages can vary significantly from object to object. 
The metallicity of stars is mostly low, and the metal enrichment during the
galaxy life looks insignificant.

A number of common relations follow from our results:

\begin{itemize}

\item The stellar mass range of the studied galaxies is more than the order of 
magnitude from about $10^6$ to $2.5\cdot 10^7$. There is a obvious relation
between the measured total stellar mass and total absolute stellar magnitude 
of the studied dwarf galaxies, in the sense that more
luminous galaxy has higher stellar mass

\item According to our measurements, dwarf galaxies in our sample of higher
luminosity forms the larger part of the stars ($\geq$60 per cent) during the initial
outburst of star formation ($\geq 12$ Gyr ago). One interesting exception is
the Tucana dSph galaxy, which possibly experienced an interaction with the Andromeda
and/or other dwarfs during an expected close proximity in the central part of
the Local galaxy Group

\item We've considered the residual star formation in dwarf spheroidal galaxies
depending on the degree of isolation of the object. Highly isolated dwarfs clearly
show recent residual star formation ($>10$ per cent of the total stellar mass), while within
the virial radius it is practically zero. Tucana is again out of the trend, located among the galaxies
inside the virial radius. It also could be an argument in favour that in the past
this dwarf was among the closer satellites of M\,31.

\end{itemize}

\section*{Acknowledgements}
The work was performed within the SAO RAS state assignment in the part
`Conducting Fundamental Science Research'.

\section*{Data availability}
The data underlying this article are available in the article and in its online supplementary material.

\bibliographystyle{mnras}
\bibliography{sfh_textv1}

\begin{thebibliography}{}
\makeatletter
\relax
\def\mn@urlcharsother{\let\do\@makeother \do\$\do\&\do\#\do\^\do\_\do\%\do\~}
\def\mn@doi{\begingroup\mn@urlcharsother \@ifnextchar [ {\mn@doi@}
  {\mn@doi@[]}}
\def\mn@doi@[#1]#2{\def\@tempa{#1}\ifx\@tempa\@empty \href
  {http://dx.doi.org/#2} {doi:#2}\else \href {http://dx.doi.org/#2} {#1}\fi
  \endgroup}
\def\mn@eprint#1#2{\mn@eprint@#1:#2::\@nil}
\def\mn@eprint@arXiv#1{\href {http://arxiv.org/abs/#1} {{\tt arXiv:#1}}}
\def\mn@eprint@dblp#1{\href {http://dblp.uni-trier.de/rec/bibtex/#1.xml}
  {dblp:#1}}
\def\mn@eprint@#1:#2:#3:#4\@nil{\def\@tempa {#1}\def\@tempb {#2}\def\@tempc
  {#3}\ifx \@tempc \@empty \let \@tempc \@tempb \let \@tempb \@tempa \fi \ifx
  \@tempb \@empty \def\@tempb {arXiv}\fi \@ifundefined
  {mn@eprint@\@tempb}{\@tempb:\@tempc}{\expandafter \expandafter \csname
  mn@eprint@\@tempb\endcsname \expandafter{\@tempc}}}

\bibitem[\protect\citeauthoryear{{Bullock} \& {Boylan-Kolchin}}{{Bullock} \&
  {Boylan-Kolchin}}{2017}]{bullock2017}
{Bullock} J.~S.,  {Boylan-Kolchin} M.,  2017, \mn@doi [\araa]
  {10.1146/annurev-astro-091916-055313}, \href
  {https://ui.adsabs.harvard.edu/abs/2017ARA&A..55..343B} {55, 343}

\bibitem[\protect\citeauthoryear{{Dolphin}}{{Dolphin}}{2000}]{dolphin2000}
{Dolphin} A.~E.,  2000, \mn@doi [\pasp] {10.1086/316630}, \href
  {https://ui.adsabs.harvard.edu/abs/2000PASP..112.1383D} {112, 1383}

\bibitem[\protect\citeauthoryear{{Dressler}}{{Dressler}}{1980}]{dressler1980}
{Dressler} A.,  1980, \mn@doi [\apj] {10.1086/157753}, \href
  {http://cdsads.u-strasbg.fr/abs/1980ApJ...236..351D} {236, 351}

\bibitem[\protect\citeauthoryear{{Garrison-Kimmel} et~al.,}{{Garrison-Kimmel}
  et~al.}{2019}]{garrison2019}
{Garrison-Kimmel} S.,  et~al., 2019, \mn@doi [\mnras] {10.1093/mnras/stz2507},
  \href {https://ui.adsabs.harvard.edu/abs/2019MNRAS.489.4574G} {489, 4574}

\bibitem[\protect\citeauthoryear{{Grebel}}{{Grebel}}{2001}]{grebel2001}
{Grebel} E.~K.,  2001, \mn@doi [Astrophysics and Space Science Supplement]
  {10.1023/A:1012742903265}, \href
  {https://ui.adsabs.harvard.edu/abs/2001ApSSS.277..231G} {277, 231}

\bibitem[\protect\citeauthoryear{{Karachentsev}, {Kashibadze}, {Makarov}  \&
  {Tully}}{{Karachentsev} et~al.}{2009}]{karach2009}
{Karachentsev} I.~D.,  {Kashibadze} O.~G.,  {Makarov} D.~I.,   {Tully} R.~B.,
  2009, \mn@doi [\mnras] {10.1111/j.1365-2966.2008.14300.x}, \href
  {https://ui.adsabs.harvard.edu/abs/2009MNRAS.393.1265K} {393, 1265}

\bibitem[\protect\citeauthoryear{{Karachentsev}, {Makarov}  \&
  {Kaisina}}{{Karachentsev} et~al.}{2013}]{upgc2013}
{Karachentsev} I.~D.,  {Makarov} D.~I.,   {Kaisina} E.~I.,  2013, \mn@doi [\aj]
  {10.1088/0004-6256/145/4/101}, \href
  {https://ui.adsabs.harvard.edu/abs/2013AJ....145..101K} {145, 101}

\bibitem[\protect\citeauthoryear{{Karachentsev}, {Makarova}, {Tully}, {Wu}  \&
  {Kniazev}}{{Karachentsev} et~al.}{2014}]{kar2014}
{Karachentsev} I.~D.,  {Makarova} L.~N.,  {Tully} R.~B.,  {Wu} P.-F.,
  {Kniazev} A.~Y.,  2014, \mn@doi [\mnras] {10.1093/mnras/stu1217}, \href
  {http://cdsads.u-strasbg.fr/abs/2014MNRAS.443.1281K} {443, 1281}

\bibitem[\protect\citeauthoryear{{Karachentsev}, {Makarova}, {Makarov}, {Tully}
   \& {Rizzi}}{{Karachentsev} et~al.}{2015}]{kar2015}
{Karachentsev} I.~D.,  {Makarova} L.~N.,  {Makarov} D.~I.,  {Tully} R.~B.,
  {Rizzi} L.,  2015, \mn@doi [\mnras] {10.1093/mnrasl/slu181}, \href
  {http://cdsads.u-strasbg.fr/abs/2015MNRAS.447L..85K} {447, L85}

\bibitem[\protect\citeauthoryear{{Kirby}, {Gilbert}, {Escala}, {Wojno},
  {Guhathakurta}, {Majewski}  \& {Beaton}}{{Kirby} et~al.}{2020}]{kirby2020}
{Kirby} E.~N.,  {Gilbert} K.~M.,  {Escala} I.,  {Wojno} J.,  {Guhathakurta} P.,
   {Majewski} S.~R.,   {Beaton} R.~L.,  2020, \mn@doi [\aj]
  {10.3847/1538-3881/ab5f0f}, \href
  {https://ui.adsabs.harvard.edu/abs/2020AJ....159...46K} {159, 46}

\bibitem[\protect\citeauthoryear{{Klypin}, {Zhao}  \& {Somerville}}{{Klypin}
  et~al.}{2002}]{klypin2002}
{Klypin} A.,  {Zhao} H.,   {Somerville} R.~S.,  2002, \mn@doi [\apj]
  {10.1086/340656}, \href
  {https://ui.adsabs.harvard.edu/abs/2002ApJ...573..597K} {573, 597}

\bibitem[\protect\citeauthoryear{{Makarov} \& {Makarova}}{{Makarov} \&
  {Makarova}}{2004}]{mm2004}
{Makarov} D.~I.,  {Makarova} L.~N.,  2004, \mn@doi [Astrophysics]
  {10.1023/B:ASYS.0000031838.50078.1a}, \href
  {http://cdsads.u-strasbg.fr/abs/2004Ap.....47..229M} {47, 229}

\bibitem[\protect\citeauthoryear{{Makarov}, {Makarova}, {Sharina}, {Uklein},
  {Tikhonov}, {Guhathakurta}, {Kirby}  \& {Terekhova}}{{Makarov}
  et~al.}{2012}]{mak2012}
{Makarov} D.,  {Makarova} L.,  {Sharina} M.,  {Uklein} R.,  {Tikhonov} A.,
  {Guhathakurta} P.,  {Kirby} E.,   {Terekhova} N.,  2012, \mn@doi [\mnras]
  {10.1111/j.1365-2966.2012.21581.x}, \href
  {http://cdsads.u-strasbg.fr/abs/2012MNRAS.425..709M} {425, 709}

\bibitem[\protect\citeauthoryear{{Makarova} et~al.,}{{Makarova}
  et~al.}{2002}]{mak2002}
{Makarova} L.~N.,  et~al., 2002, \mn@doi [\aap] {10.1051/0004-6361:20021426},
  \href {http://cdsads.u-strasbg.fr/abs/2002A%26A...396..473M} {396, 473}

\bibitem[\protect\citeauthoryear{{Makarova}, {Makarov}, {Karachentsev}, {Tully}
   \& {Rizzi}}{{Makarova} et~al.}{2017}]{mak2017}
{Makarova} L.~N.,  {Makarov} D.~I.,  {Karachentsev} I.~D.,  {Tully} R.~B.,
  {Rizzi} L.,  2017, \mn@doi [\mnras] {10.1093/mnras/stw2502}, \href
  {http://cdsads.u-strasbg.fr/abs/2017MNRAS.464.2281M} {464, 2281}

\bibitem[\protect\citeauthoryear{{Martin} et~al.,}{{Martin}
  et~al.}{2017}]{martin2017}
{Martin} N.~F.,  et~al., 2017, \mn@doi [\apj] {10.3847/1538-4357/aa901a}, \href
  {https://ui.adsabs.harvard.edu/abs/2017ApJ...850...16M} {850, 16}

\bibitem[\protect\citeauthoryear{{Monelli} et~al.,}{{Monelli}
  et~al.}{2010}]{monelli2010}
{Monelli} M.,  et~al., 2010, \mn@doi [\apj] {10.1088/0004-637X/722/2/1864},
  \href {http://cdsads.u-strasbg.fr/abs/2010ApJ...722.1864M} {722, 1864}

\bibitem[\protect\citeauthoryear{{Oemler}}{{Oemler}}{1974}]{oemler1974}
{Oemler} Jr. A.,  1974, \mn@doi [\apj] {10.1086/153216}, \href
  {http://cdsads.u-strasbg.fr/abs/1974ApJ...194....1O} {194, 1}

\bibitem[\protect\citeauthoryear{{Rocha}, {Peter}  \& {Bullock}}{{Rocha}
  et~al.}{2012}]{rocha2012}
{Rocha} M.,  {Peter} A. H.~G.,   {Bullock} J.,  2012, \mn@doi [\mnras]
  {10.1111/j.1365-2966.2012.21432.x}, \href
  {https://ui.adsabs.harvard.edu/abs/2012MNRAS.425..231R} {425, 231}

\bibitem[\protect\citeauthoryear{{Saviane}, {Held}  \& {Piotto}}{{Saviane}
  et~al.}{1996}]{saviane1996}
{Saviane} I.,  {Held} E.~V.,   {Piotto} G.,  1996, \aap, \href
  {https://ui.adsabs.harvard.edu/abs/1996A&A...315...40S} {315, 40}

\bibitem[\protect\citeauthoryear{{Savino}, {Tolstoy}, {Salaris}, {Monelli}  \&
  {de Boer}}{{Savino} et~al.}{2019}]{savino2019}
{Savino} A.,  {Tolstoy} E.,  {Salaris} M.,  {Monelli} M.,   {de Boer} T.~J.~L.,
   2019, \mn@doi [\aap] {10.1051/0004-6361/201936077}, \href
  {https://ui.adsabs.harvard.edu/abs/2019A&A...630A.116S} {630, A116}

\bibitem[\protect\citeauthoryear{{Schlafly} \& {Finkbeiner}}{{Schlafly} \&
  {Finkbeiner}}{2011}]{schlafly}
{Schlafly} E.~F.,  {Finkbeiner} D.~P.,  2011, \mn@doi [\apj]
  {10.1088/0004-637X/737/2/103}, \href
  {http://cdsads.u-strasbg.fr/abs/2011ApJ...737..103S} {737, 103}

\bibitem[\protect\citeauthoryear{{Simpson}, {Grand}, {G{\'o}mez}, {Marinacci},
  {Pakmor}, {Springel}, {Campbell}  \& {Frenk}}{{Simpson}
  et~al.}{2018}]{simpson2017}
{Simpson} C.~M.,  {Grand} R. J.~J.,  {G{\'o}mez} F.~A.,  {Marinacci} F.,
  {Pakmor} R.,  {Springel} V.,  {Campbell} D. J.~R.,   {Frenk} C.~S.,  2018,
  \mn@doi [\mnras] {10.1093/mnras/sty774}, \href
  {https://ui.adsabs.harvard.edu/abs/2018MNRAS.478..548S} {478, 548}

\bibitem[\protect\citeauthoryear{{Skillman} et~al.,}{{Skillman}
  et~al.}{2017}]{skillman2017}
{Skillman} E.~D.,  et~al., 2017, \mn@doi [\apj] {10.3847/1538-4357/aa60c5},
  \href {https://ui.adsabs.harvard.edu/abs/2017ApJ...837..102S} {837, 102}

\bibitem[\protect\citeauthoryear{{Teyssier}, {Johnston}  \&
  {Kuhlen}}{{Teyssier} et~al.}{2012}]{teyssier2012}
{Teyssier} M.,  {Johnston} K.~V.,   {Kuhlen} M.,  2012, \mn@doi [\mnras]
  {10.1111/j.1365-2966.2012.21793.x}, \href
  {https://ui.adsabs.harvard.edu/abs/2012MNRAS.426.1808T} {426, 1808}

\bibitem[\protect\citeauthoryear{{Tolstoy}, {Hill}  \& {Tosi}}{{Tolstoy}
  et~al.}{2009}]{tolstoy2009}
{Tolstoy} E.,  {Hill} V.,   {Tosi} M.,  2009, \mn@doi [\araa]
  {10.1146/annurev-astro-082708-101650}, \href
  {https://ui.adsabs.harvard.edu/abs/2009ARA&A..47..371T} {47, 371}

\bibitem[\protect\citeauthoryear{{Weisz} et~al.,}{{Weisz}
  et~al.}{2011}]{weisz2011}
{Weisz} D.~R.,  et~al., 2011, \mn@doi [\apj] {10.1088/0004-637X/739/1/5}, \href
  {https://ui.adsabs.harvard.edu/abs/2011ApJ...739....5W} {739, 5}

\bibitem[\protect\citeauthoryear{{Weisz}, {Dolphin}, {Skillman}, {Holtzman},
  {Gilbert}, {Dalcanton}  \& {Williams}}{{Weisz} et~al.}{2014}]{weisz2014}
{Weisz} D.~R.,  {Dolphin} A.~E.,  {Skillman} E.~D.,  {Holtzman} J.,  {Gilbert}
  K.~M.,  {Dalcanton} J.~J.,   {Williams} B.~F.,  2014, \mn@doi [\apj]
  {10.1088/0004-637X/789/2/147}, \href
  {https://ui.adsabs.harvard.edu/abs/2014ApJ...789..147W} {789, 147}

\bibitem[\protect\citeauthoryear{{Weisz} et~al.,}{{Weisz}
  et~al.}{2019}]{weisz2019}
{Weisz} D.~R.,  et~al., 2019, \mn@doi [\apjl] {10.3847/2041-8213/ab4b52}, \href
  {https://ui.adsabs.harvard.edu/abs/2019ApJ...885L...8W} {885, L8}

\bibitem[\protect\citeauthoryear{{Williams} et~al.,}{{Williams}
  et~al.}{2017}]{wil2017}
{Williams} B.~F.,  et~al., 2017, \mn@doi [\apj] {10.3847/1538-4357/aa862a},
  \href {http://cdsads.u-strasbg.fr/abs/2017ApJ...846..145W} {846, 145}

\makeatother
\end{thebibliography}

\label{lastpage}
\end{document}